\documentclass[letter,12pt]{article}
\usepackage{amsmath,amsthm,amssymb,graphicx,natbib,setspace,dcolumn,float,booktabs,bm}
\usepackage{footmisc}
\usepackage{mathtools}
\newcolumntype{.}{D{.}{.}{-1}}
\usepackage[top = 1 in, bottom = 1 in, outer = 1 in, inner = 1 in]{geometry}
\usepackage{hyperref}
\usepackage{subcaption}

\usepackage{lscape}

\title{Reclustering:\\
A New Method to Test the Appropriate Level of Clustering
\thanks{Paper prepared for delivery at the 121st Annual Meeting of the American Political Science Association, Vancouver, September 11--14, 2025. 
}
}

\author{
Kentaro Fukumoto\thanks{Professor, Graduate Schools for Law and Politics, The University of Tokyo. Mailing address:~7-3-1 Hongo, Bunkyo, Tokyo 113-0033, Japan. 
Email address:~\href{mailto:fukumoto@j.u-tokyo.ac.jp}{\texttt{fukumoto@j.u-tokyo.ac.jp}}. 
ORCiD: 0000-0003-3704-9054. 
}
}

\begin{document}

\maketitle

\begin{abstract}
When scholars suspect units are dependent on each other within clusters but independent of each other across clusters, they employ cluster-robust standard errors (CRSEs). Nevertheless, what to cluster over is sometimes unknown. For instance, in the case of cross-sectional survey samples, clusters may be households, municipalities, counties, or states. A few approaches have been proposed, although they are based on asymptotics. I propose a new method to address this issue that works in a finite sample: reclustering. That is, we randomly and repeatedly group fine clusters into new gross clusters and calculate a statistic such as CRSEs. Under the null hypothesis that fine clusters are independent of each other, how they are grouped into gross clusters should not matter for any cluster-sensitive statistic. Thus, if the statistic based on the original clustering is a significant outlier against the distributions of the statistics induced by reclustering, it is reasonable to reject the null hypothesis and employ gross clusters. I compare the performance of reclustering with that of a few previous tests using Monte Carlo simulation and application.

\vspace{10pt}

\end{abstract}

\newpage
\doublespacing

\section*{INTRODUCTION\centering}

When scholars suspect that units are dependent on each other within clusters but independent of each other across clusters, they employ cluster-robust standard errors (CRSEs) \citep{Angrist_Pischke_2009, Cameron_Miller_2015, MacKinnon_Nielsen_Webb_2023_guide}. The suspicion is, however, not necessarily validated in many empirical analyses. Moreover, when there are a few nested levels for clustering, scholars usually choose one of them without any formal test. For instance, in the case of survey samples, clusters may be household, municipality, county, or state. Another example is experiments in schools, where researchers may cluster over class room or school. Dependence may occur due to spatial spillover or temporal persistence.

As I elaborate on later, a few approaches have been proposed, although they are based on asymptotics. I propose a new method to address this issue that works in a finite sample: reclustering. That is, we randomly and repeatedly group fine clusters into new gross clusters and calculate a 
statistics such as CRSEs. Under the null hypothesis that fine clusters are independent of each other, how we divide fine clusters into gross clusters should not matter for any 
gross cluster-sensitive statistic. Thus, if the statistic based on the original gross clustering is a significant outlier against the distributions of the statistics induced by reclustering, it is reasonable to reject the null hypothesis and employ CRSE. 
Reclustering is a new method to derive the distribution of a test statistic for the appropriate level of clustering. 

The paper is organized as follows. The next section lays out the problem. The following section presents reclustering in a simple case so that readers easily understand its core idea. In the fourth section, I generalize the setup and elaborate on reclustering. In the fifth section, I consider the performance of reclustering and the few previous tests by Monte Carlo simulation. In the penultimate section, I compare these methods by applying them to a famous study  about the effect of foot ball game win on incumbent's electoral win \citep{GHMM}. 
Finally, I conclude.

\section*{AGENDA\centering}

\subsection*{Cluster Robust Standard Error}

Suppose there are $n$ units, and each unit is indexed by $i \in \{1,2,\ldots,n\}$. 
For a while, I consider one level of clustering.
All units are divided into $\bar{g}$ clusters, 
and each cluster is indexed by $g \in \{1,2,\ldots,\bar{g}\}$. 
The cluster to which unit $i$ belongs is denoted by $g[i]$.

The $n$-dimension (column) vector of the dependent variable is denoted by $\bm y$. 
Supposing there are $\bar{k}$ independent variables, the $n \times \bar{k}$ matrix of the independent variables is denoted by $\bm X$. The regression equation is
\begin{equation*}
\bm y = \bm X' \bm \beta + \bm u,
\end{equation*}
where $\bm \beta$ is the $\bar{k}$-dimension vector of the coefficients, and $\bm u$ is the $n$-dimension vector of the error term. 
The OLS coefficient estimator of $\bm \beta$ is $\hat{\bm \beta} = (\bm X' \bm X)^{-1}\bm X'\bm y = \bm \beta + (\bm X' \bm X)^{-1} \bm s$, where the $\bar{k}$-dimension vector 
$\bm s \equiv \bm X' \bm u = \sum_{i = 1}^{n} \bm s_{i}$ is the sum of score vectors, $\bm s_{i} \equiv \bm x_{i} u_{i}$, 
and $\bm x_{i}$ is the column vector that corresponds to the $i$-th row of $\bm X$. 
The conventional estimator of the variance covariance matrix of $\hat{\bm \beta}$ is
\begin{equation*}
\hat{V}(\hat{\bm \beta}) \equiv 
(\bm X' \bm X)^{-1} \hat{\bm \Sigma} (\bm X' \bm X)^{-1},
\end{equation*}
where $\hat{\bm \Sigma}$ is an estimator of $\bm \Sigma \equiv  E(\bm s \bm s') = \sum_{i = 1}^{n} \sum_{i' = 1}^{n} E(\bm s_{i} \bm s'_{i'})$. 

Here I consider two hypotheses about $\bm \Sigma$. 
\begin{description}
\item [Null Hypothesis 1] All units are independent of each other.
\item [Alternative Hypothesis 1] Units in different clusters are independent of each other, while some units in a cluster are not.
\end{description}
Let the $\bar{k} \times \bar{k}$ square matrix all elements of which are equal to zero be denoted by $\bm 0$. 
Under Null Hypothesis 1, $E(\bm s_{i} \bm s'_{i'}) = \bm 0$ in the case of $i \neq i'$, and thus $\bm \Sigma$ is reduced to $\bm \Sigma^{\mathrm{naive}} \equiv \sum_{i = 1}^{n} E(\bm s_{i} \bm s'_{i})$. 
Under Alternative Hypothesis 1, 
$E(\bm s_{i} \bm s'_{i'}) = \bm 0$ in the case of $g[i] \neq g[i']$, and for some $i \neq i'$ such that $g[i] = g[i']$, $E(\bm s_{i} \bm s'_{i'}) \neq \bm 0$, and thus $\bm \Sigma$ is only reduced to $\bm \Sigma^{\mathrm{CR}} \equiv \sum_{g = 1}^{\bar{g}} E(\bm s_{g} \bm s'_{g})$, where $\bm s_{g} \equiv \sum_{i: g[i] = g} \bm s_{i}$.

Accordingly, we estimate $\bm \Sigma^{\mathrm{naive}}$ and $\bm \Sigma^{\mathrm{CR}}$ by 
\begin{equation*}
\begin{split}
\hat{\bm \Sigma}^{\mathrm{naive}} & \equiv \sum_{i = 1}^{n} \hat{\bm s}_{i} \hat{\bm s}'_{i} \\
\hat{\bm \Sigma}^{\mathrm{CR}} & \equiv \sum_{g = 1}^{\bar{g}} \hat{\bm s}_{g} \hat{\bm s}'_{g},
\end{split}
\end{equation*}
respectively, where $\hat{\bm s}_{i}$ and $\hat{\bm s}_{g}$ are estimators of  $\bm s_{i}$ and $\bm s_{g}$, respectively. Below, I focus on $\hat{\bm s}_{i} \equiv \frac{n}{n-\bar{k}} \bm x_{i} \hat{u}_{i}$ (so called HC1) and $\hat{\bm s}_{g} \equiv \frac{\bar{g}(n - 1)}{(\bar{g}-1)n} \sum_{i: g[i] = g} \hat{\bm s}_{i}$ (so called CV1), where $\hat{u}_{i} \equiv y_{i} - \bm x_{i}' \hat{\bm \beta}$, although the argument below holds for other estimators of $\bm s_{i}$ and $\bm s_{g}$ such as so called HC2/CV2 and HC3/CV3 \citep{MacKinnon_Nielsen_Webb_2023_guide}. 
Then, CRSE is
\begin{equation*}
\begin{split}
\mathrm{CRSE}(\hat{\bm \beta})
&\equiv \sqrt{\hat{V}^{\mathrm{CR}} (\hat{\bm \beta}) }\\
&\equiv \sqrt{(\bm X' \bm X)^{-1} \hat{\bm \Sigma}^{\mathrm{CR}} (\bm X' \bm X)^{-1}}.
\end{split}
\end{equation*}

\subsection*{Problem}

Even under Null Hypothesis 1, CRSE is still valid. 
Nonetheless, it is less efficient than the conventional standard error because CRSE takes into account off diagonal cross products of scores $
 \hat{\bm s}_{i} \hat{\bm s}'_{i}$ ($i \neq i$), which should be equal to zero on average and thus the conventional standard error ignores. In a sample, these cross products are not exactly equal to zero. 
As \citet[132]{Hansen_2022} notes, it is possible that ``clustering adds sampling noise and is thus less accurate.''

On the other hand, 
under Alternative Hypothesis 1, the conventional standard error is not valid. 
When the number of clusters $\bar{g}$ is not large enough, however, asymptotic approximation does not work well, CRSE suffers from sampling error, and the confidence intervals tend to under-cover \citep{Angrist_Pischke_2009}.

The central problem is that we do not know which is true, Null Hypothesis 1 
or Alternative Hypothesis 1. To address the problem, scholars usually follow a few textbook 
rule of thumb. 
According to \citet[333]{Cameron_Miller_2015}, ``[t]he consensus is to be conservative ... 
and to use bigger and more aggregate clusters when possible, up to and including the point at which there is concern about having too few clusters.''
\citet[321]{Angrist_Pischke_2009} recommend, ``a conservative approach is ... take the larger of robust and conventional standard errors.''

These ``conservative'' guidelines are, however, not based on any theory. Moreover, if Null Hypothesis 1 is true, CRSE may be unnecessarily larger than the conventional standard error. 
In addition, since there is no clear-cut definition of ``few'' \citep[319]{Cameron_Miller_2015}, students do not know when they stop persuing ``bigger and more aggregate clusters.'' 

Lately, \citet[33]{Abadie_Athey_Imbens_Wooldridge_2023} present a new framework in the context of causal inference, arguing ``[i]f sampling is not clustered, standard errors should be clustered at the treatment assignment level'' and ``[w]hen sampling and assignment are random, clustering standard errors is not appropriate.''

\subsection*{Previous Tests
}

Recently, a few tests of the appropriate level of clustering are proposed. 
Below, I focus on one ($k$-th) explanatory variable $x_{k, i}$ as well as the corresponding coefficient $\beta_{k}$'s standard error and score $s_{k, i} \equiv x_{k, i} u_{i}$.\footnote{It is easy to partial out variables to reduce a multiple regression to a simple regression.} 

First, \citet{MacKinnon_Nielsen_Webb_2023_SV} propose the score variance (SV) test, where their test statistic $\tau^{\mathrm{SV}}$ standardizes $\sum_{g = 1}^{\bar{g}}(\sum_{i: g[i] = g} \hat{s}_{k, i})^2 - \sum_{i = 1}^{n} \hat{s}_{k, i}^2$, which is the $k$-th diagonal element of $\hat{\bm \Sigma}^{\mathrm{CR}} - \hat{\bm \Sigma}^{\mathrm{naive}}$ 
and should be close to zero under Null Hypothesis 1.
\footnote{\citet{MacKinnon_Nielsen_Webb_2023_SV} also propose another SV test for not just one but all explanatory variables.
}

Second, \citet{Ibragimov_Muller_2016} estimate $\hat{\beta}_{k, g}$ for each cluster $g$, and their test statistic is an estimate of the \underline{v}ariance of the (scaled) \underline{m}ean of $\hat{\beta}_{k, g}$'s: $\tau^{\mathrm{VMB}} \equiv \frac{n}{\bar{g}(\bar{g}-1)} \sum_{g = 1}^{\bar{g}} (\hat{\beta}_{k, g} - \bar{\hat{\beta}}_{k})^2$ where $\bar{\hat{\beta}}_{g} \equiv \frac{1}{\bar{g}} \sum_{g = 1}^{\bar{g}} \hat{\beta}_{k, g}$. 

Finally, \citet{Cai_2024} introduce the worst-case randomization (WCR) test. 
The test statistic $\tau^{\mathrm{WCR}}$ conservatively estimates an average net number of positive scores which is, in essence, $\frac{1}{\bar{g}} \sum_{g = 1}^{\bar{g}} | \sum_{i: g[i] = g}\{ I(s_{k, i} > 0) - I(s_{k, i} < 0)\} |$. 
It should be close to zero under Null Hypothesis 1. 
The SV and WCR tests 
exploits the spirit of Hausman test: even if Null Hypothesis 1 holds, CRSE is still valid. 

\section*{SIMPLE CASE\centering}

\subsection*{Method}

I introduce reclustering. 
To be simple, suppose 

each cluster has $n_{G} \equiv n / \bar{g}$ units. 

We permute clustering $\bar{r}$ times; that is, we permute the order of units and assign each unit to the cluster of the corresponding position. Specifically, in the $r$-th round, denote a new (permuted) order of unit $i$ by $i^{(r)}[i]$. Then, a new cluster of unit $i$ is $g^{(r)}[i] \equiv g[i^{(r)}[i]]$. Thus, each new cluster has $n_{G}$ units again. 

I call this procedure reclustering.
Note that this is not bootstrap; we do not resample with replacement. 

A test statistic 
should satisfy only two requirements. 
Define $\bm D = (\bm y, \bm X)$ and $\bm g \equiv (g[1], g[2], \ldots, g[n])'$.
Let the test statistic denoted by $\tau(\bm D, \bm g)$, a function of $\bm D$ and $\bm g$. 
First, the test statistic should be row-exchangeable; $\tau(\bm D^{(r)}, \bm g^{(r)}) = \tau(\bm D, \bm g)$ where $\bm D^{(r)}$ represents the matrix whose $i$-th row is $\bm d^{\prime(r)}_{i} \equiv \bm d'_{i^{(r)}[i]}$, $\bm d_{i} = (y_{i}, \bm x'_{i})'$, and $\bm g^{(r)} \equiv (g^{(r)}[1], g^{(r)}[2], \ldots, g^{(r)}[n])'$. 
Second, the test statistic should be ``cluster-sensitive'' in the sense that it is not degenerated into a function of only $\bm D$; for some (usually, most) $r$, $\tau(\bm D, \bm g) \neq \tau(\bm D, \bm g^{(r)})$.\footnote{As the probability $\tau(\bm D, \bm g^{(r)}) \neq \tau(\bm D, \bm g^{(r')})$ is higher in the case of $r \neq r'$, the $p$-value of the test statistic (which I expalin shortly) becomes more fine-grained.
} 
We can use the test statistics proposed by prior research, while the typical one I think of is the CRSE of $\beta_{k}$: $\tau^{\mathrm{CRSE}}(\bm D, \bm g) \equiv \sqrt{\hat{V}^{\mathrm{CR}}_{kk}(\hat{\bm \beta})}$, the square root of the $k$-th row and the $k$-th column element of $\hat{V}^{\mathrm{CR}}(\hat{\bm \beta})$. 
Since $\hat{\bm \beta}$ is a function of $\bm y$ and $\bm X$ but not $\bm g$, $\hat{\bm \beta}$ is not cluster-sensitive, and I can write $\hat{\bm \beta}(\bm D)$.
Accordingly, the point estimate $\hat{\bm \beta}(\bm D)$ remains the same through reclustering, and thus $\hat{\bm \beta}(\bm D)$ cannot be $\tau(\bm D, \bm g)$ even though 
$\hat{\bm \beta}(\bm D)$ is row-exchangeable: $\hat{\bm \beta}(\bm D^{(r)}) = \hat{\bm \beta}(\bm D)$.

We calculate the test statistic for each set of reclusters, $\tau(\bm D, \bm g^{(r)})$, where we do not permute $\bm D$. For instance, if $\bm X$ includes cluster dummy variables, we continue to use dummy variables to indicate the original cluster ($I(g[i] = g)$), not those to indicate permuted recluster ($I(g^{(r)}[i] = g)$). 
I denote the observed cluster vector by $\bm g^{\mathrm{obs}}$ and calculate the corresponding test statistic $\tau(\bm D, \bm g^{\mathrm{obs}})$. Then, we calculate the following $p$-value:
\begin{equation}
p \equiv \frac{1}{\bar{r}}
\sum_{r = 1}^{\bar{r}} I\{ \tau(\bm D, \bm g^{(r)}) > \tau(\bm D, \bm g^{\mathrm{obs}})\}. \label{main}
\end{equation}
Let significance level $\alpha$. 
When we conduct two-side test, 
we reject the null hypothesis in the cases of $p < \alpha / 2$ or $p \geq 1 - \alpha / 2$. 
When we conduct one-side test, 
we reject the null hypothesis in the case of $p < \alpha$ (or $p \geq 1 - \alpha$).

\subsection*{Justification}

Before I formally justify reclustering, I do so informally and intuitively. 
Under Null Hypothesis 1, clusters and reclusters should not matter for the variance covariance matrix of $\hat{\bm \beta}$. 
For instance, under Null Hypothesis 1, $\tau^{\mathrm{CRSE}}(\bm D, \bm g^{(r)})$ is as valid a CRSE as $\mathrm{CRSE}(\hat{\bm \beta})_{k}$. 
Therefore, $\tau(\bm D, \bm g^{(r)})$'s should not be so different from $\tau(\bm D, \bm g^{\mathrm{obs}})$. Nevertheless, if $\tau(\bm D, \bm g^{\mathrm{obs}})$ is a significant outlier against the distribution of $\tau(\bm D, \bm g^{(r)})$'s, we should doubt Null Hypothesis 1 but support Alternative Hypothesis 1 and employ CRSE.

Below, I formalize my argument. 
Denote the distributions of $\bm d_{i}$ and $\bm D$ by $p_{d}(\bm d_{i})$ and $p_{D}(\bm D)$, respectively. 
I assume $\bm g$ is fixed.
Under Null Hypothesis 1, 

\begin{equation*}
\begin{split}
p_{D}(\bm D) 
&= \prod_{i=1}^{n} p_{d}(\bm d_{i})  \quad (\because \textrm{Null Hypothesis 1})\\
&= \prod_{i^{(r)}=1}^{n} p_{d}(\bm d_{i^{(r)}}) \quad (\because \textrm{permutation})\\
&=p_{D}(\bm D^{(r)}) \quad (\because \textrm{Null Hypothesis 1})
\end{split}
\end{equation*}
This equation implies that $\bm D$ and $\bm D^{(r)}$ can be regarded as random samples from the same population. 
Two remarks follow. 
For ease of exposition, 
I assume $\bm d_{i} \neq \bm d_{i'}$ if $i \neq i'$. 
Define $\mathcal{D}$ as the set of all permutations of $\bm D$. Note that $|\mathcal{D}| = n!$. 
First, $p_{D}(\bm D^{(r)})$ is constant irrespective of $r$. Thus, the probability of $\bm D^{(r)}$ conditioned on $\bm D^{(r)} \in \mathcal{D}$ is $1/|\mathcal{D}|$. 
Second, 

the distribution of $\tau(\bm D^{(r)}, \bm g)$ is equivalent to the distribution of $\tau(\bm D, \bm g)$, and thus, the probability of $\tau(\bm D, \bm g)$ conditioned on $\bm D \in \mathcal{D}$ is also $1/|\mathcal{D}|$. 
Thus, the conditional probability is
\begin{equation}
\begin{split}
p_{T}(t \mid \mathcal{D}) 
&\equiv \Pr\{\tau(\bm D, \bm g) > t \mid \bm D \in \mathcal{D}\} \\
&= \frac{1}{|\mathcal{D}|} \sum_{\bm D \in \mathcal{D}} I\{ \tau(\bm D, \bm g) > t \}. \label{conditional}
\end{split}
\end{equation}
We can approximate it by
\begin{equation}
\begin{split}
\hat{p}_{T}(t \mid \mathcal{D}) 
&\equiv \frac{1}{\bar{r}} \sum_{r = 1}^{\bar{r}} I\{ \tau(\bm D^{(-r)}, \bm g) > t \}\\
&= \frac{1}{\bar{r}} \sum_{r = 1}^{\bar{r}} I\{ \tau(\bm D, \bm g^{(r)}) > t \} \quad (\because \textrm{row-exchangeability}), 
\label{approximation} 
\end{split}
\end{equation}
where $\bm D^{(-r)}$ represents the matrix whose $i$-th row is $\bm d^{\prime(-r)}_{i} \equiv \bm d'_{i^{(-r)}[i]}$ such that $i^{(r)}[i^{(-r)}[i]] = i$, and it holds $\tau(\bm D^{(-r)}, \bm g) = \tau(\bm D, \bm g^{(r)})$ thanks to row-exchangeability. 
Thus, the unconditional probability is
\begin{equation}
\begin{split}
p_{T}(t) 
&\equiv \Pr\{\tau(\bm D, \bm g) > t \}\\
&= \int \Pr\{\tau(\bm D, \bm g) > t \mid \bm D \in \mathcal{D}\} p(\mathcal{D}) 
d \mathcal{D}. \label{unconditional}
\end{split}
\end{equation}
If $\bm D$ is a random sample, $\hat{p}_{T}(t \mid \mathcal{D}) $ is an unbiased estimator of $p_{T}(t)$. I substitute $t = \tau(\bm D, \bm g^{\mathrm{obs}})$. 

Note that we do not need a large value of $\bar{g}$ or $n$; thus, reclustering does not rely on asymptotics. For good approximation of Equation (\ref{conditional}) by Equation (\ref{approximation}), my experience suggests that $\bar{r}=1,000$ is sufficiently large.

\section*{GENERAL CASE\centering}

\subsection*{Problem}

I extend the simple case in a few directions. First, we have two levels of clusters, gross (or coarse) and fine (or subcluster), and fine clusters are nested within a gross cluster. Second, gross (fine) clusters have various numbers of fine clusters (individual units).

Now I introduce new notation. 
Suppose there are $n$ units, and each unit is indexed by $i \in \{1,2,\ldots,n\}$. 
All units are divided into $\bar{g}$ \underline{g}ross clusters, 
and each gross cluster is indexed by $g \in \{1,2,\ldots,\bar{g}\}$ and has $n_{g}$ \underline{f}ine clusters. Across gross clusters, fine cluster is indexed by $f \in \{1,2,\ldots,\bar{f}\}$ and has $n_{f}$ units such that $\bar{f} \equiv \sum_{g = 1}^{\bar{g}} n_{g}$ and $\sum_{f = 1}^{\bar{f}} n_{f} = n$.
The fine cluster to which unit $i$ belongs is denoted by $f[i]$. 
The gross cluster to which fine cluster $f$ belongs is denoted by $g[f]$. 
Note that when $n_{f} = 1$ for all $f$, the fine cluster is essentially a unit, and the setup is reduced to one where there is only one level cluster. Furthermore, when $n_{g}$ is constantly equalt to $n_{G}$ for all $g$, the setup is reduced to the simple case.

Here I consider two new hypotheses about $\bm \Sigma$. 
\begin{description}
\item [Null Hypothesis 2] Units in different fine clusters are independent of each other, while some units in a fine cluster are not.
\item [Alternative Hypothesis 2] Units in different gross clusters are independent of each other, while some units in a gross cluster are not.
\end{description}
Define $\bm s_{f} \equiv \sum_{i: f[i] = f} \bm s_{i}$ and redefine $\bm s_{g} \equiv \sum_{f: g[f] = g} \bm s_{f}$. 
Under Null Hypothesis 2, $E(\bm s_{f} \bm s'_{f'}) = \bm 0$ in the case of $f \neq f'$, and thus $\bm \Sigma$ is reduced to $\bm \Sigma^{\mathrm{naive}} \equiv \sum_{f = 1}^{\bar{f}} E(\bm s_{f} \bm s'_{f})$. 
Under Alternative Hypothesis 2, $E(\bm s_{f} \bm s'_{f'}) = \bm 0$ in the case of $g[f] \neq g[f']$, and for some $f \neq f'$ such that $g[f] = g[f']$, $E(\bm s_{f} \bm s'_{f'}) \neq \bm 0$, and thus $\bm \Sigma$ is only reduced to $\bm \Sigma^{\mathrm{CR}} \equiv \sum_{g = 1}^{\bar{g}} E(\bm s_{g} \bm s'_{g})$.

Thus, CRSE at gross cluster level takes into account fine cluster level off diagonal cross products of scores $\hat{\bm s}_{f} \hat{\bm s}'_{f'}$ ($f \neq f'$), which should be equal to zero on average under Null Hypothesis 2 and thus CRSE at fine cluster level ignores. In a sample, these cross products are not exactly equal to zero even under Null Hypothesis 2 due to sampling error. 

\subsection*{Previous Tests
}

The three previous tests are adapted to the general case in the following way.
First, the test statistic of the SV test standardizes $\sum_{g = 1}^{\bar{g}}(\sum_{i: g[i] = g} \hat{s}_{k, i})^2 - \sum_{f = 1}^{\bar{f}}(\sum_{i: f[i] = f} \hat{s}_{k, i})^2$\citep{MacKinnon_Nielsen_Webb_2023_SV}.
Second, the test statistic of the VMB test remains the same except where the standard error of $\hat{\beta}_{k, g}$ is clustered at fine cluster level \citep{Ibragimov_Muller_2016}. 
Finally, the test statistic of the WCR test conservatively estimates, in essence, $\frac{1}{\bar{g}} \sum_{g = 1}^{\bar{g}} | \sum_{f: g[f] = g}\{ I(s_{k, f} > 0) - I(s_{k, f} < 0)\} |$, where $s_{k, f} \equiv \sum_{i: f[i] = f}s_{k, i}$ \citep{Cai_2024}. 

The distributions of these three latest test statistics are derived based on asymptotics. According to \citet[938]{Cai_2024}, the SV, VMB, and WCR tests require large values of $\bar{g}, n_g$ and $n_f$, respectively.\footnote{\citet{MacKinnon_Nielsen_Webb_2023_SV} consider a large value of $n$, while they admit that a fixed value of $\bar{g}$ has not been well addressed (in their Section 4.4). They also dismiss finite sample corrections, which are dependent on $\bar{g}$ and $\bar{f}$, implicitly assuming large values of $\bar{g}$ and $\bar{f}$. However, it does not matter when one use cluster wild bootstrap.

\citet[90]{Ibragimov_Muller_2016} consider a large value of $n_g$.}

Therefore, their performance is not (at least, theoretically) guaranteed when the sample size is small.

\subsection*{Reclustering
}

We permute the order of fine clusters and assign each fine cluster to the gross cluster of the corresponding position. Specifically, in the $r$-th round, denote a new (permuted) order of fine cluster $f$ by $f^{(r)}[f]$. Then, a new gross cluster of fine cluster $f$ is $g^{(r)}[f] \equiv g[f^{(r)}[f]]$. Thus, new gross cluster $g^{(r)} = g$ has $n_{g}$ fine clusters again.\footnote{Although the number of units in new gross cluster $g^{(r)} = g$, $\sum_{f: g^{(r)}[f] =g} n_{f}$, may change, it should not matter. This is because most test statistics take into account fine level information such as $\bm s_{f} \equiv \sum_{i: f[i] =f} \bm s_{i}$, not unit level information such as $\bm s_{i}$.} 
Note that the fine cluster to which unit $i$ belongs, $f[i]$, does not change through reclustering. That is, we retain fine-level cluster structure,  but not gross level one. 

Define $\bm D_{f}$ 
as the matrix of stacked $(y_{i}, \bm x'_{i}, f[i])$'s where $f[i] = f$ and $\bm g [f]$ is a $n_f$ dimension vector all of whose elements are $g [f]$. 
Redefine $\bm D$ as the matrix of stacked $\bm D_{f}$'s: $\bm D \equiv (\bm D'_{1}, \bm D'_{2}, \ldots \bm D'_{\bar{f}})'$ and $\bm g \equiv (\bm g'[1], \bm g'[2], \ldots, \bm g'[\bar{f}])'$. 
A test statistic $\tau(\bm D, \bm g)$ should satisfy only two requirements.  
First, it should be ``fine-cluster-exchangeable;'' $\tau(\bm D^{(r)}, \bm g^{(r)}) = \tau(\bm D, \bm g)$ where $\bm D^{(r)} \equiv (\bm D'_{f^{(r)}[1]}, \bm D'_{f^{(r)}[2]}, \ldots \bm D'_{f^{(r)}[\bar{f}]})'$ 
and $\bm g^{(r)} \equiv (\bm g^{\prime(r)}[1], \bm g^{\prime(r)}[2], \ldots, \bm g^{\prime(r)}[\bar{f}])'$ where $\bm g^{(r)} [f]$ is a $n_f$ dimension vector all of whose elements are $g^{(r)} [f]$. 

Second, it should be ``gross-cluster-sensitive'' in the sense that it is not degenerated into a function of only $\bm D$.

Then, we can calculate the $p$-value in the same way as the simple case, using Equation \ref{main}.

\subsection*{Justification}

Denote the distributions of $\bm D_{f}$ and $\bm D$ by $p_{d}(\bm D_{f} \mid n_{f})$ and $p_{D}(\bm D)$, respectively. 
Under Null Hypothesis 2,
\begin{equation*}
\begin{split}
p_{D}(\bm D) 
&= \prod_{f=1}^{\bar{f}} p_{d}(\bm D_{f} \mid n_{f})  \quad (\because \textrm{Null Hypothesis 2})\\
&= \prod_{f^{(r)}=1}^{\bar{f}} p_{d}(\bm D_{f^{(r)}} \mid n_{f^{(r)}}) \quad (\because \textrm{permutation})\\
&=p_{D}(\bm D^{(r)}) \quad (\because \textrm{Null Hypothesis 2})
\end{split}
\end{equation*}
This equation implies that $\bm D$ and $\bm D^{(r)}$ can be regarded as random samples from the same population. It holds $\tau(\bm D^{(-r)}, \bm g) = \tau(\bm D, \bm g^{(r)})$ thanks to fine-cluster-exchangeability. 
I assume $\bm g$ is fixed.
Thus, like the simple case, Equations (\ref{conditional}) to (\ref{unconditional}) hold, and $\hat{p}_{T}(t \mid \mathcal{D}) $ is an unbiased estimator of $p_{T}(t)$.

\subsection*{Very Small Sample}

When the significance level is $\alpha$ and we use a two-tailed test, we need at least $2 / \alpha$ different values of $\tau(\bm D, \bm g^{(r)})$, namely, different partitions of fine clusters into gross clusters. The number of ways we divide $\bar{f}$ fine clusters into $\bar{g}$ gross clusters, which is denoted by $\bar{r}^{*}$, is
\begin{equation*}
\bar{r}^{*} \equiv \frac{ \bar{f}! }{ \bar{g}! \prod_{g = 1}^{\bar{g}} n_{g}! }.
\end{equation*}
Table \ref{tab:bar_r_star} presents the value of $\bar{r}^{*}$ for very small values of $\bar{g}$ and $n_{G} = \bar{f} / \bar{g}$. Suppose $\alpha = 0.05$. Thus, $\bar{r}^{*}$ should be not smaller than $2 / 0.05 = 40$. When $\bar{g} = 2$, we need at least $n_{G} \geq 5$.  When $\bar{g} = 3$, we need at least $n_{G} \geq 3$. When $\bar{g} \geq 4$, we do not have to worry about $n_{G} $. 

\begin{table}[!htb]
\centering
\caption{The number of ways we divide $\bar{f}$ fine clusters into $\bar{g}$ gross clusters.}
\label{tab:bar_r_star}
\begin{tabular}{rrrr}
  \hline
&\multicolumn{3}{l}{$\bar{g}$}\\
$n_{G}$ & 2 & 3 & 4 \\ 
  \hline
2 & 3 & 15 & 105 \\ 
  3 & 10 & 280 & 15,400 \\ 
  4 & 35 & 5,775 & 2,627,625 \\ 
  5 & 126 & 126,126 & 488,864,376 \\ 
   \hline
\end{tabular}

\end{table}

\section*{SIMULATION\centering}

In the simulation, I compare the performance of the four tests I have considered so far. 
Following prior research \citep{Cai_2024, Ibragimov_Muller_2016,  MacKinnon_Nielsen_Webb_2023_SV}, 
I set the following scenario. 
For ease of presentation, I change the notation of the index. Denote fine cluster $f$ in gross cluster $g$ by $gf$. Index $f$ counts from one to $n_{g}$ in each gross cluster $g$. Denote unit $i$ in fine cluster $gf$ by $gfi$. Index $i$ counts from one to $n_{gf}$ in each fine cluster $gf$. 

\subsection*{Method}

\subsubsection*{Clustered variable generation model}

I make the generic clustered variable $q_{gfi}$ by either a first-order autoregressive (AR1) model or a hidden factor model so that even after controlling for cluster fixed effects, some correlation among units in a cluster remains. 

I begin with an AR1 model \citep{Cai_2024, Ibragimov_Muller_2016}. 

In the case of a fine-level clustered variable, I draw a white noise $\epsilon_{gfi}$ from the standard normal distribution for each unit $gfi$. 
Let $q_{gf1} = \epsilon_{gf1}$. 
For $i \geq 2 
$, let $q_{gfi} = \rho q_{gf, i-1} +\sqrt{1 - \rho^2} \epsilon_{gfi}$, where $0 \leq \rho < 1$ is a pre-decided correlation parameter. Note that 
$E(q_{gfi}q_{gfi'}) = \rho^{|i-i'|}$. 
In the case of a gross-level clustered variable, 
Let $q_{g11} = \epsilon_{g11}$. 
For $i \geq 2 
$, let $q_{gfi} = \rho q_{gf, i-1} +\sqrt{1 - \rho^2} \epsilon_{gfi}$.
For $f \geq 2$, let $q_{gf1} = \rho q_{g, (f-1), n_{g, (f-1)}} +\sqrt{1 - \rho^2} \epsilon_{gf1}$. 
I randomly reorder $q_{gfi}$ in each gross cluster so that some units far apart across different fine clusters within the same gross cluster are sufficiently correlated with each other, when $E(q_{gfi} q_{gf'i'})$ is sufficiently large even if $f \neq f'$.\footnote{Even if I do not reorder units, the smallest correlation $E(q_{g11} q_{gn_{g}n_{gn_{g}}}) = \rho^{\sum_{gf= g1}^{gn_{g}} n_{gf}}$
 is not exactly equal to zero but too small for any test to detect.
}

Next, I explain a hidden factor model \citep{MacKinnon_Nielsen_Webb_2023_SV}.
In the case of a fine-level clustered variable, I sample factors $\epsilon_{gf}^{\mathrm{odd}}$ and $\epsilon_{gf}^{\mathrm{even}}$ from the standard normal distribution for each fine cluster $gf$. 
I also sample a white noise $\epsilon_{gfi}$ from the standard normal distribution for each unit $gfi$. 
I make $q_{gfi} = \rho \epsilon_{gf}^{\mathrm{odd}} +\sqrt{1 - \rho^2} \epsilon_{gfi}$ if $i$ is odd and $q_{gfi} = \rho \epsilon_{gf}^{\mathrm{even}} +\sqrt{1 - \rho^2} \epsilon_{gfi}$ if $i$ is even. Note that 
$E(q_{gfi}q_{gfi'}) = \rho^2$ if $i \equiv i' \pmod 2$ and $E(q_{gfi}q_{gfi'}) = 0$ otherwise.
In the case of a gross-level clustered variable, I replace $\epsilon_{gf}^{\mathrm{odd}}$ and $\epsilon_{gf}^{\mathrm{even}}$ by $\epsilon_{g}^{\mathrm{odd}}$ and $\epsilon_{g}^{\mathrm{even}}$, and I sample them from the standard normal distribution for each gross cluster $g$.

Note that in both models, each $q_{gfi}$ follows the standard normal distribution, independently of each other across clusters (thus $E(q_{gfi}q_{g'f'i'}) = 0$ if $gf \neq g'f'$ or $g \neq g'$) but not (necessarily in the case of the hidden factor model) within a cluster. 

\subsubsection*{Iteration}

In one iteration of a simulation, I conduct the procedure below. 
According to the hidden factor model 
mentioned above, I make the gross-level clustered explanatory variable $x^{G}_{gfi}$ and error term $u^{G}_{gfi}$ with $\rho = \rho^{G}_{X}$ and $\rho = \rho^{G}_{U}$, respectively.\footnote{I randomly reorder $x^{G}_{gfi}$ and $u^{G}_{gfi}$ in the same way so that units within the same gross cluster are correlated with each other with respect to not only $x^{G}_{gfi}$ and $u^{G}_{gfi}$ but also $
s^{G}_{gfi} \equiv x^{G}_{gfi} u^{G}_{gfi}$.
I did not use the AR1 model simply because I did not have much time; in a new version, I will use it.}
I also make fine-level clustered variables $x^{F}_{gfi}$ and $u^{F}_{gfi}$ with $\rho = \rho^{F}_{X}$ and $\rho = \rho^{F}_{U}$, respectively. 
I make $x_{gfi} \equiv 0.5 
x^{F}_{gfi} + 0.5 
x^{G}_{gfi}$ and $u_{gfi} \equiv 0.5 
 u^{F}_{gfi} +  0.5 
u^{G}_{gfi}$, 
so that in the case of $
\rho^{F}_{X} 
\rho^{F}_{U} \neq 0$, $x_{gfi}$ and $u_{gfi}$ are correlated with each other within a fine cluster even if $\rho^{G}_{X} = 0$ or $\rho^{G}_{U} = 0$. 
Note that $x_{gfi}$ and $u_{gfi}$ follow the standard normal distribution. 

I also independently draw $\bar{f}$ fine cluster fixed effects, $\phi_{f}$'s, from the standard normal distribution. 
Then, I make
\begin{equation*}
y_{gfi} = \beta x_{gfi} + \phi_{f} + u_{gfi},
\end{equation*}
where $\beta = 1$. I include fixed effects because I would like to consider the situation where fixed effects cannot explain away within-cluster correlation.

I regress $y_{gfi}$ on $x_{gfi}$ and $\bar{f}$ dummy variables, each of which is an indicator of each fine cluster ($I(gf[i] = gf)$), without the constant term. We consider four test statistics: $\tau^{\mathrm{CRSE}}, \tau^{\mathrm{SV}}, \tau^{\mathrm{VMB}}$, and $\tau^{\mathrm{WCR}}$.\footnote{For the WCR test, I slightly adapted the author's R script at \url{https://github.com/yong-cai/WCR/} (accessed on May 5, 2025).}
I derive $p$-values for $\tau^{\mathrm{CRSE}}$ 
by reclustering, 
for $\tau^{\mathrm{SV}}$ by cluster wild bootstrap, and for $\tau^{\mathrm{VMB}}$ and $\tau^{\mathrm{WCR}}$ by Monte Carlo simulation.\footnote{Theoretically, reclustering can be used for any test statistic. Nevertheless, when I used it for the three test statistics other than CRSE, their performance was not good.
The $p$-values for $\tau^{\mathrm{SV}}$ can be also derived by asymptotics, although the performance is worse than that by cluster wild bootstrap according to \citet{MacKinnon_Nielsen_Webb_2023_SV} and my own experience.}

While all methods involve repetition in calculating $p$-values, 
I repeat 1,000 times (e.g., $\bar{r}=1,000$ for reclustering).\footnote{As for bootstrap, I resample $1,000 - 1 = 999$ times \citep{MacKinnon_Nielsen_Webb_2023_SV}.
} 

I iterate the above procedure $z$ times, which varies across setups simply due to trial and error.\footnote{I plan to conduct all simulations with $z = 10,000$.}

\subsection*{Results}

The baseline setup is $\bar{g} = n_{g} = 12, n_{gf} = 100$ for all $g$ and $f$, respectively, $\rho^{G}_{X} = \rho^{F}_{X}= 0.5$, and $\rho^{F}_{U} = \rho^{G}_{U} \equiv \rho_{U}$. 

\begin{figure}[!htb]
\centering
    \includegraphics[width=0.49\hsize]{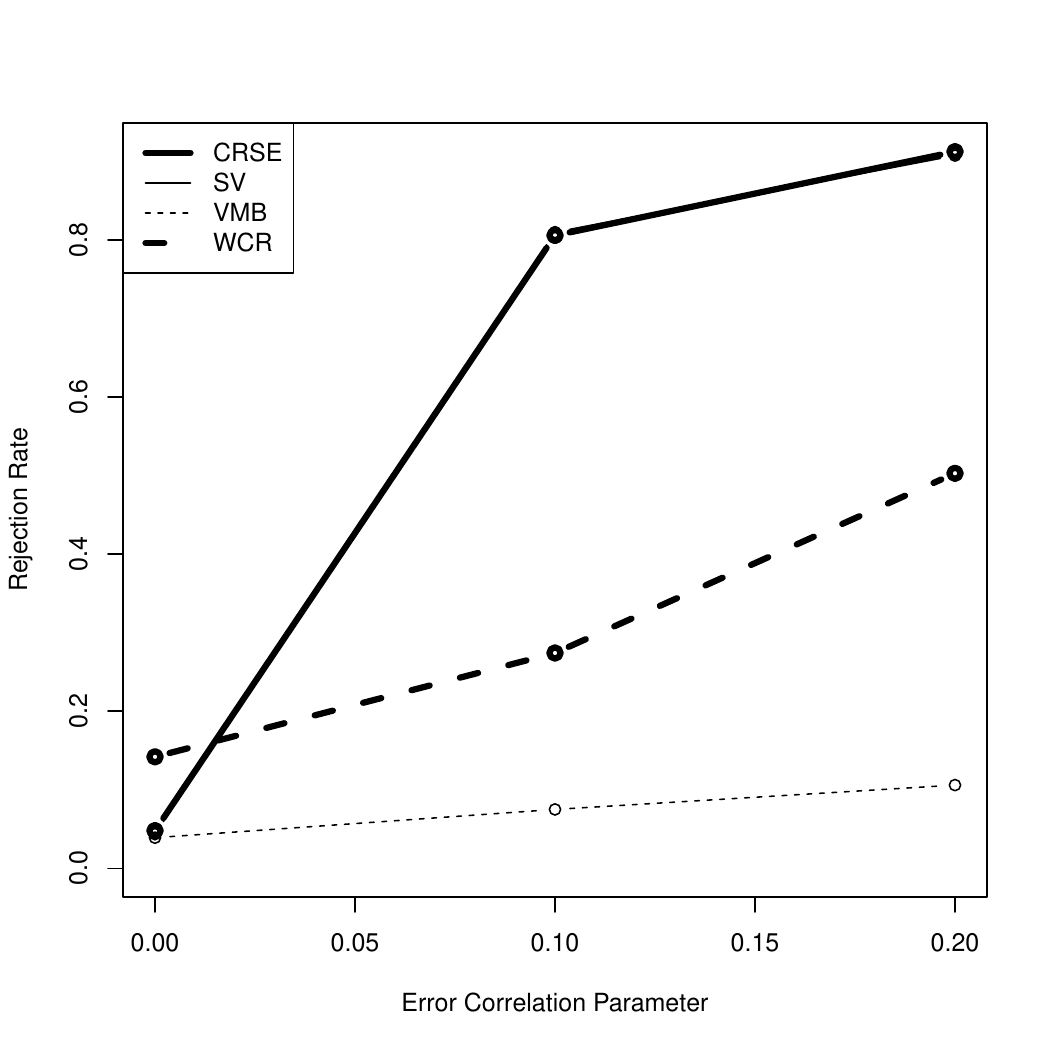}
\caption{How the error correlation parameter affects the performance of the four tests. I denote the rejection rate of the CRSE test by thick solid lines, that of the SV test by thin solid lines, that of the VMB test by thin dotted lines, and that of the WCR test by thick dotted lines. $\bar{g} = n_{g} = 12, n_{gf} = 100$ for all $g$ and $f$, respectively. $z = 2,000$ for $\rho_{U} = 0, 0.2$ and $z = 1,200$ for $\rho_{U} = 0.1$.}
\label{fig:rho}
\end{figure}

To begin, I examine the performance of the four tests as I increase the error correlation parameter $\rho_{U}$ from 0 ($z = 2,000$) through 0.1 ($z = 1,200$) to 0.2 ($z = 2,000$). Other setups are the same as the baseline. Figure \ref{fig:rho} shows the results. 
The horizontal axis indicates $\rho_{U}$, while the vertical axis represents the rejection rate, the rate to reject Null Hypothesis 2 by a two-sided test at the significance level $\alpha = 0.05$. 
I denote the rejection rate of the CRSE test by thick solid lines, that of the SV test by thin solid lines (which is almost the same as that of the CRSE test and thus is almost hidden), that of the VMB test by thin dotted lines, and that of the WCR test by thick dotted lines. 
Three remarks are in order. First, in the case of $\rho_{U} = 0$, when Null Hypothesis 2 holds, all but the WCR tests reject Null Hypothesis 2 at the nominal level, while the WCR test over-rejects it. 
Second, the power increases in $\rho_{U}$ for all tests.
Third, the VMB test almost lacks power.

\subsubsection*{Numbers of Clusters and Units}

Now I study how the numbers of clusters and units affects the performance of the four tests. When I check the size of the test, I set $
\rho_{U} = 0, z = 2,000$, where Null Hypothesis 2 is true. When I check the power of the test, I set $\rho_{U} = 0.1, z = 1,200$, where Alternative Hypothesis 2 is true.

\begin{figure}[!htb]
\centering
\begin{minipage}[b]{0.49\hsize}
    \centering
    \includegraphics[width=0.9\hsize]{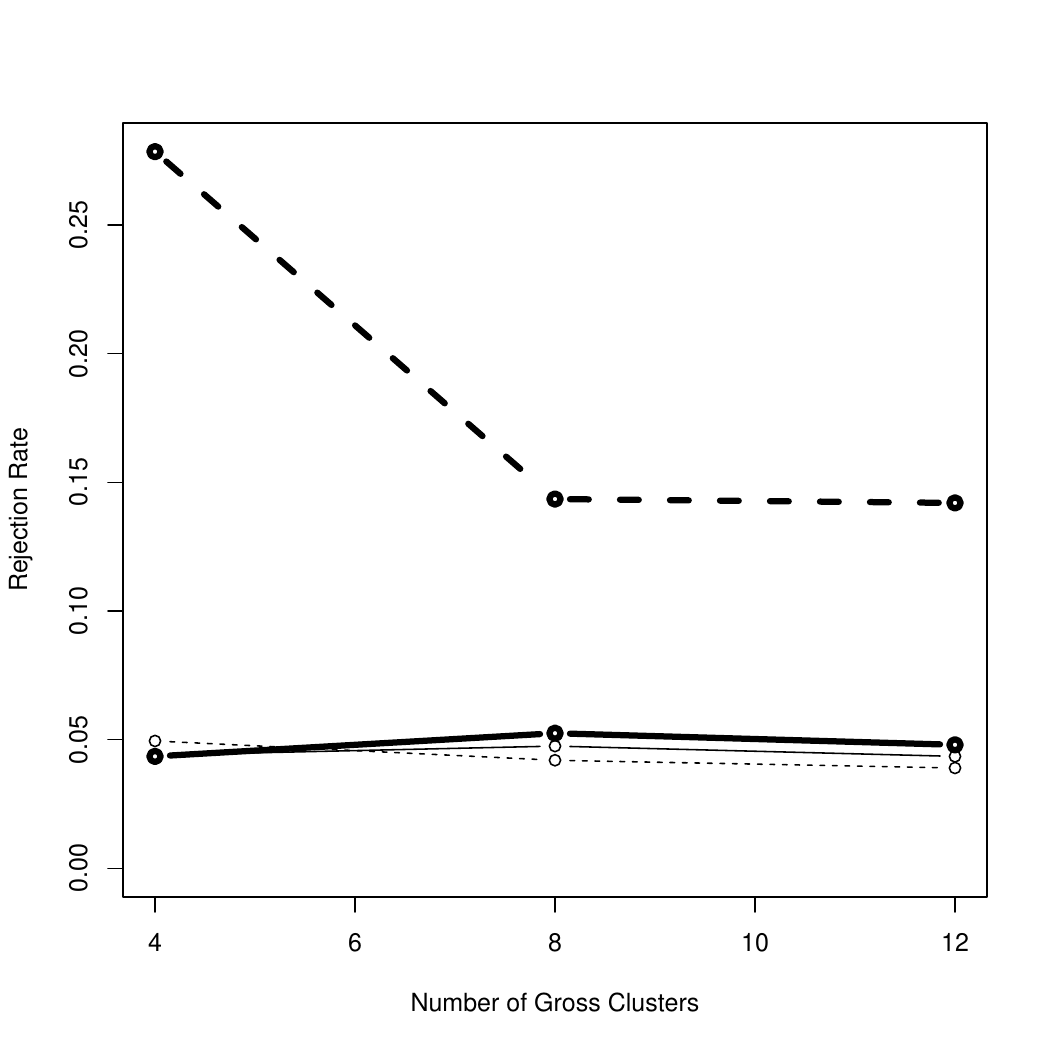}
    \subcaption{Size}
\end{minipage}
\begin{minipage}[b]{0.49\hsize}
    \centering
    \includegraphics[width=0.9\hsize]{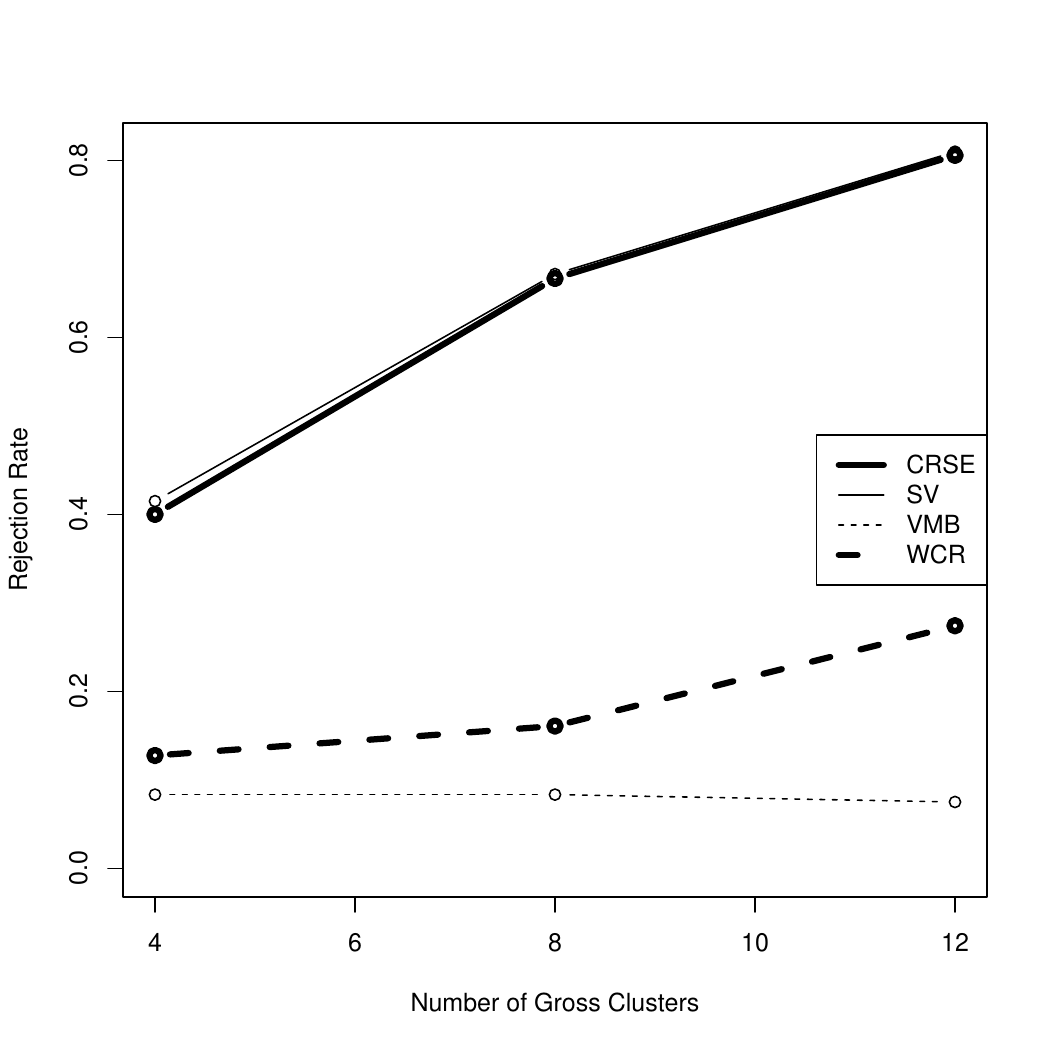}
    \subcaption{Power}
\end{minipage}
\caption{How the number of gross clusters affects the performance of the four tests. I denote the rejection rate of the CRSE test by thick solid lines, that of the SV test by thin solid lines, that of the VMB test by thin dotted lines, and that of the WCR test by thick dotted lines. $n_{g} = 12, n_{gf} = 100$ for all $g$ and $f$, respectively. In the left panel, $\rho_{U} = 0, z = 2,000$, while in the right panel, $\rho_{U} = 0.1, z = 1,200$.}
\label{fig:G}
\end{figure}

First, I examine the performance of the four tests as I change the number of gross clusters $\bar{g}$ from 4 through 8 to 12. Figure \ref{fig:G} shows the results. The left and right panels correspond to $\rho_{U} = 0$ and $\rho_{U} = 0.1$, respectively. In each panel, the horizontal axis indicates $\bar{g}$. Under Null Hypothesis 2 (left panel), the rejection rate of all but the WCR tests do not depends on $\bar{g}$, while the WCR test over-rejects Null Hypothesis 2  more frequently in the case of a small number of gross clusters ($\bar{g} = 4$). Under Alternative Hypothesis 2 (right panel), all but the VMB tests reject Null Hypothesis 2 more frequently as $\bar{g}$ becomes larger. The rejection rate of the CRSE test is almost the same as that of the SV test (and thus the latter is almost hidden by the former), although that of the WCR test is far behind. \citet[938]{Cai_2024} writes that the SV test requires a large value of $\bar{g}$, although it can detect the presence of gross clusters even in the case of a small value of $\bar{g}$.

Second, I move the number of fine clusters in each gross cluster $n_{g}$ from 4 through 8 to 12, which is constant across $g$'s. Other setups are the same as the baseline.
Figure \ref{fig:n_{g}} shows the results. 
The horizontal axis indicates $n_{g}$. 
Since it is similar to Figure \ref{fig:G}, I only mention the difference between the two below. Under Null Hypothesis 2 (left panel), the VMB test overrejects Null Hypothesis 2 in the case of a small number of fine clusters ($n_{g} = 4$). 
This result reflects that the test need a large number of $n_{g}$ \citep[938]{Cai_2024}. 
Under Alternative Hypothesis 2 (right panel), the reject rate of the VMB test is larger than 5\%; however, considering over-rejection under the null hypothesis, this may be not because the test has some power.

\begin{figure}[!htb]
\centering
\begin{minipage}[b]{0.49\hsize}
    \centering
    \includegraphics[width=0.9\hsize]{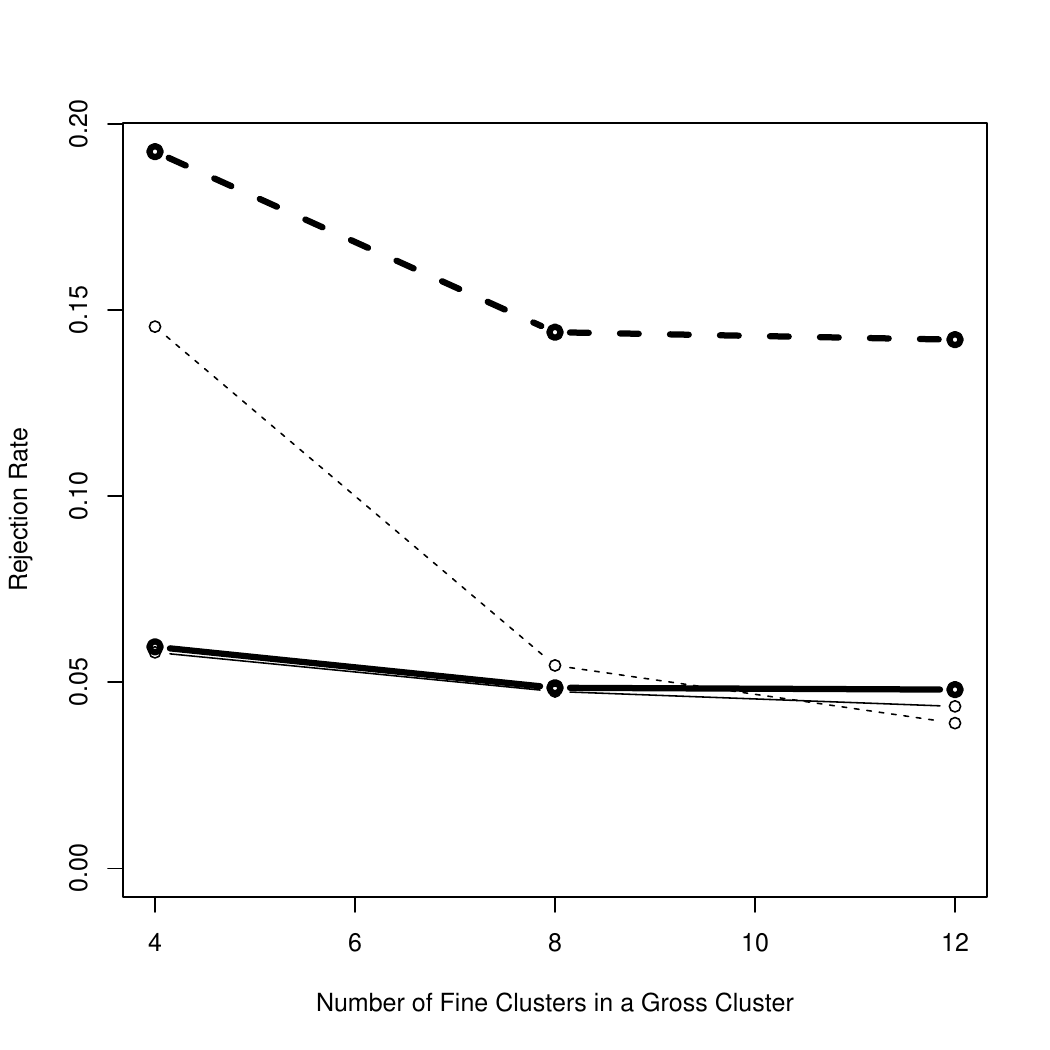}
    \subcaption{Size}
\end{minipage}
\begin{minipage}[b]{0.49\hsize}
    \centering
    \includegraphics[width=0.9\hsize]{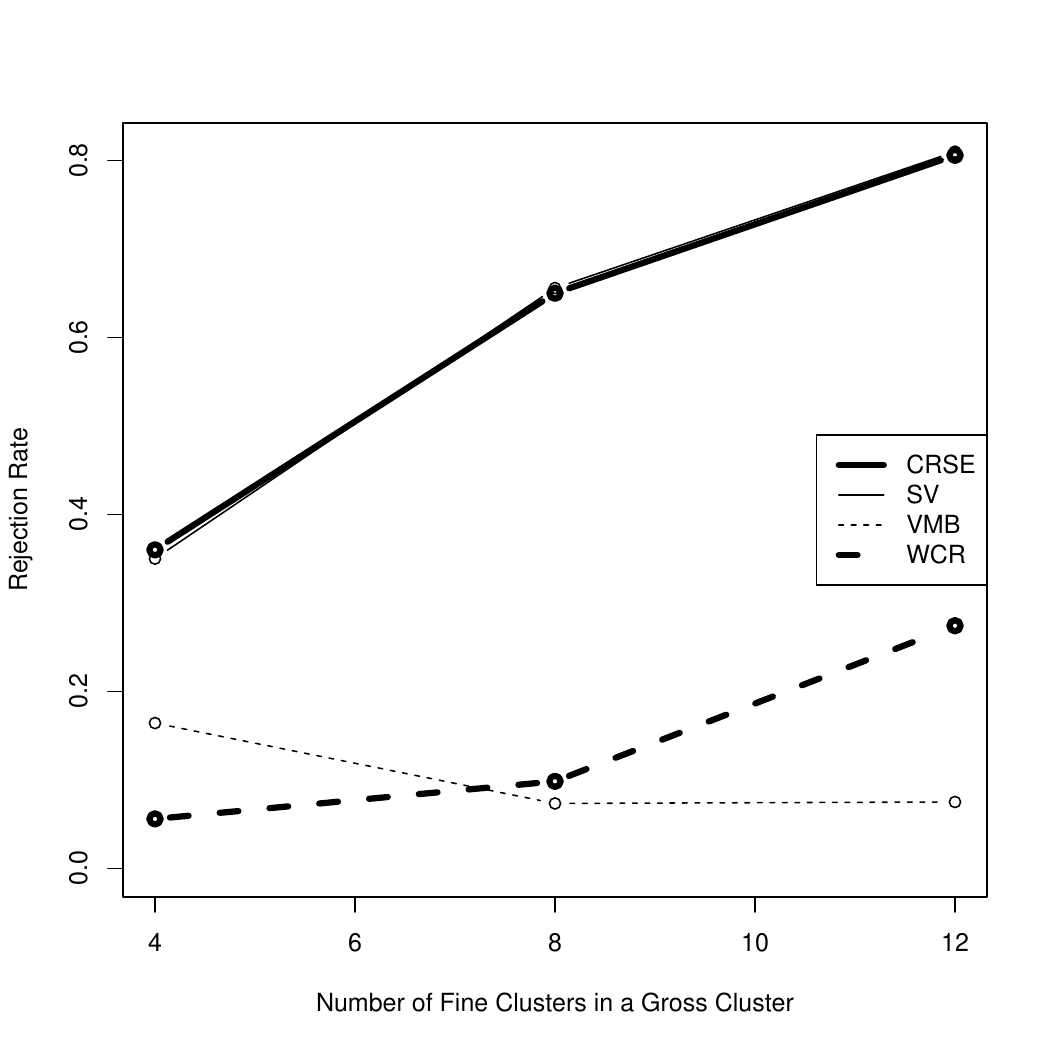}
    \subcaption{Power}
\end{minipage}
\caption{How the number of fine clusters in a gross cluster affects the performance of the four tests. I denote the rejection rate of the CRSE test by thick solid lines, that of the SV test by thin solid lines, that of the VMB test by thin dotted lines, and that of the WCR test by thick dotted lines. $\bar{g} = 12, n_{gf} = 100$ for all $g$ and $f$, respectively. In the left panel, $\rho_{U} = 0, z = 2,000$, while in the right panel, $\rho_{U} = 0.1, z = 1,200$.}
\label{fig:n_{g}}
\end{figure}

\begin{figure}[!htb]
\centering
\begin{minipage}[b]{0.49\hsize}
    \centering
    \includegraphics[width=0.9\hsize]{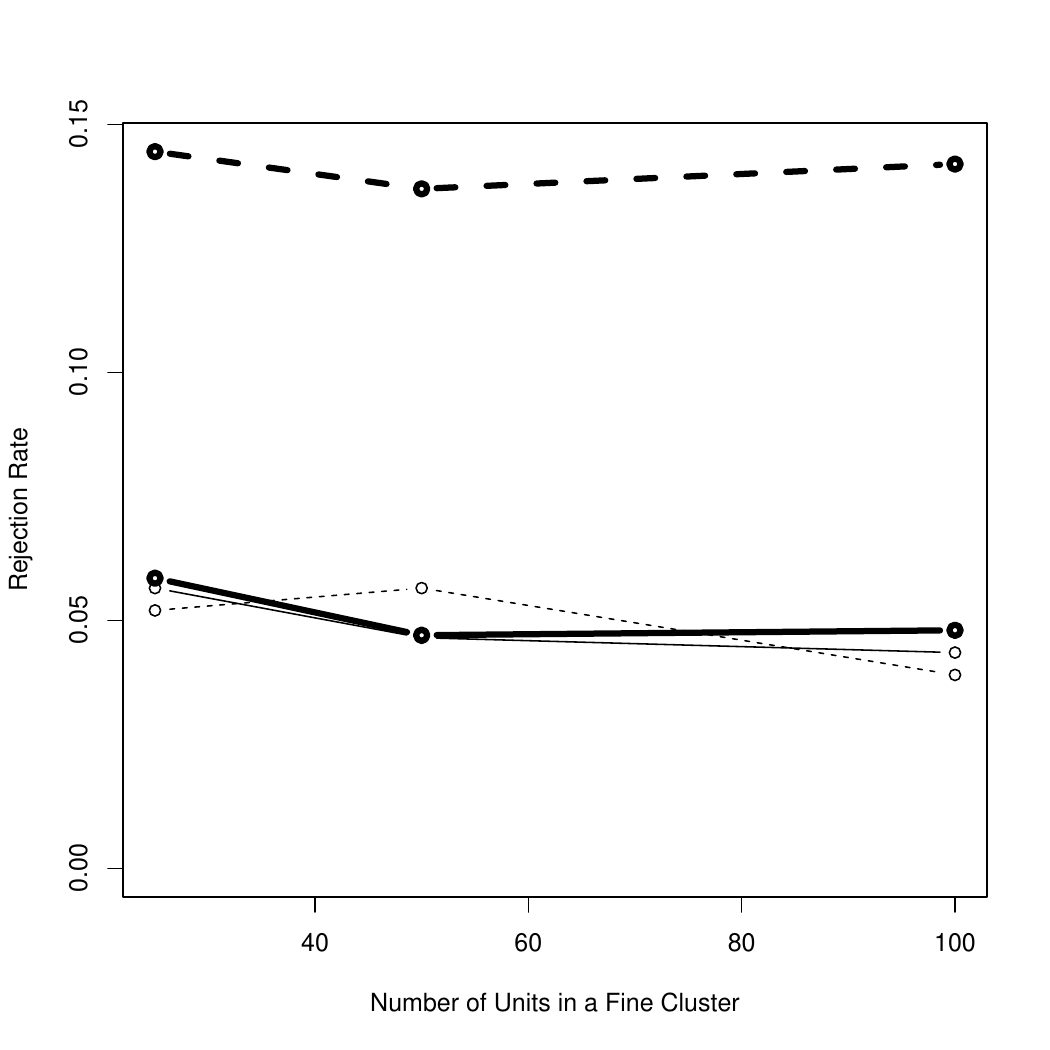}
    \subcaption{Size}
\end{minipage}
\begin{minipage}[b]{0.49\hsize}
    \centering
    \includegraphics[width=0.9\hsize]{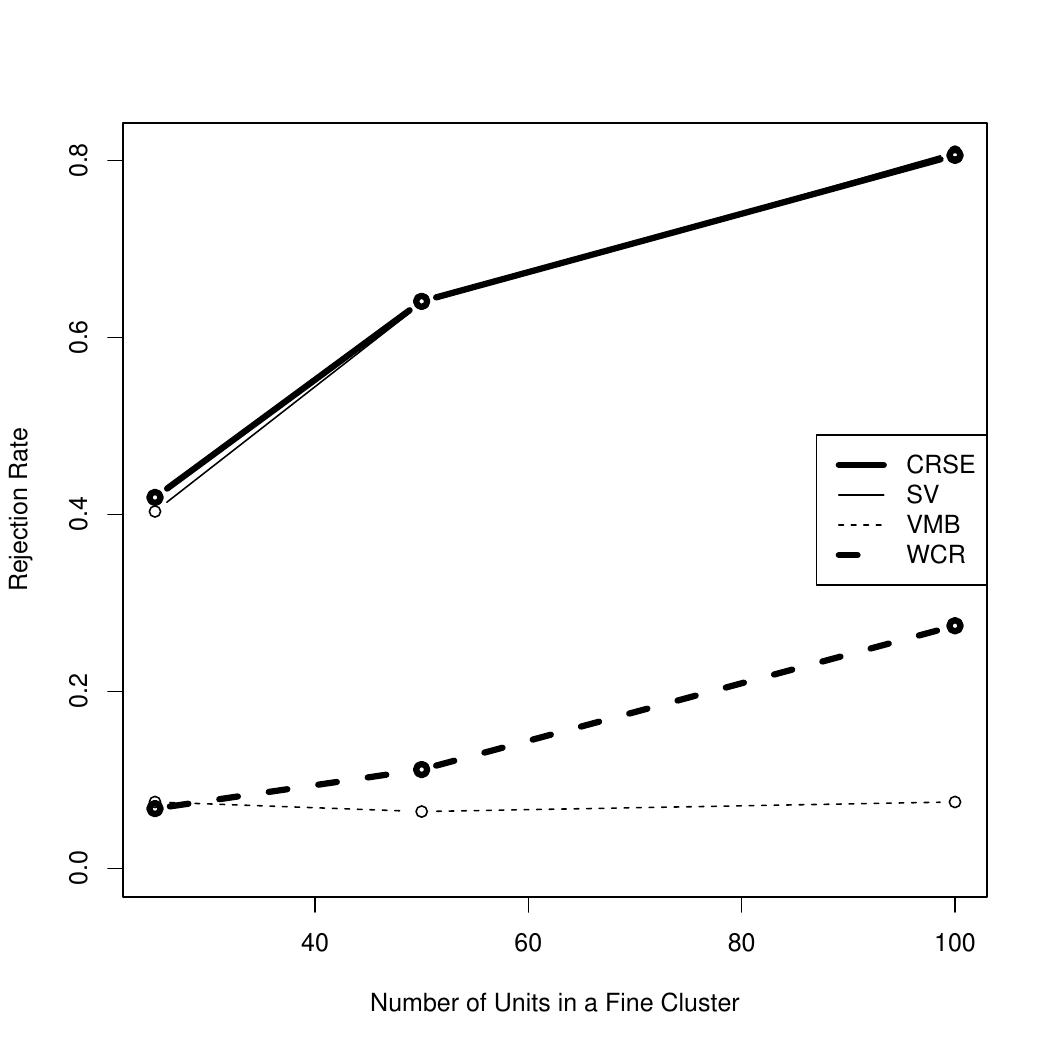}
    \subcaption{Power}
\end{minipage}
\caption{How the number of units in a fine cluster affects the performance of the four tests. I denote the rejection rate of the CRSE test by thick solid lines, that of the SV test by thin solid lines, that of the VMB test by thin dotted lines, and that of the WCR test by thick dotted lines. $\bar{g} = n_{g} = 12$ for all $g$. In the left panel, $\rho_{U} = 0, z = 2,000$, while in the right panel, $\rho_{U} = 0.1, z = 1,200$.}
\label{fig:n_f}
\end{figure}

Third, I increase the number of units in each fine cluster $n_{gf}$ from 25 through 50 to 100, which is constant across $gf$'s. Other setups are the same as the baseline.
Figure \ref{fig:n_f} shows the results. 
The horizontal axis indicates $n_{gf}$. 
It is also similar to Figure \ref{fig:G}. The only substantial difference is that the rejection rate of the WCR test is not larger in the case of a small number of units in a fine cluster ($n_{gf} = 25$). This result is unexpected because the test necessitates a large number of $n_{gf}$ \citep[938]{Cai_2024}.

\subsubsection*{Heterogeneity}

\begin{figure}[!htb]
\centering
\begin{minipage}[b]{0.49\hsize}
    \centering
    \includegraphics[width=0.9\hsize]{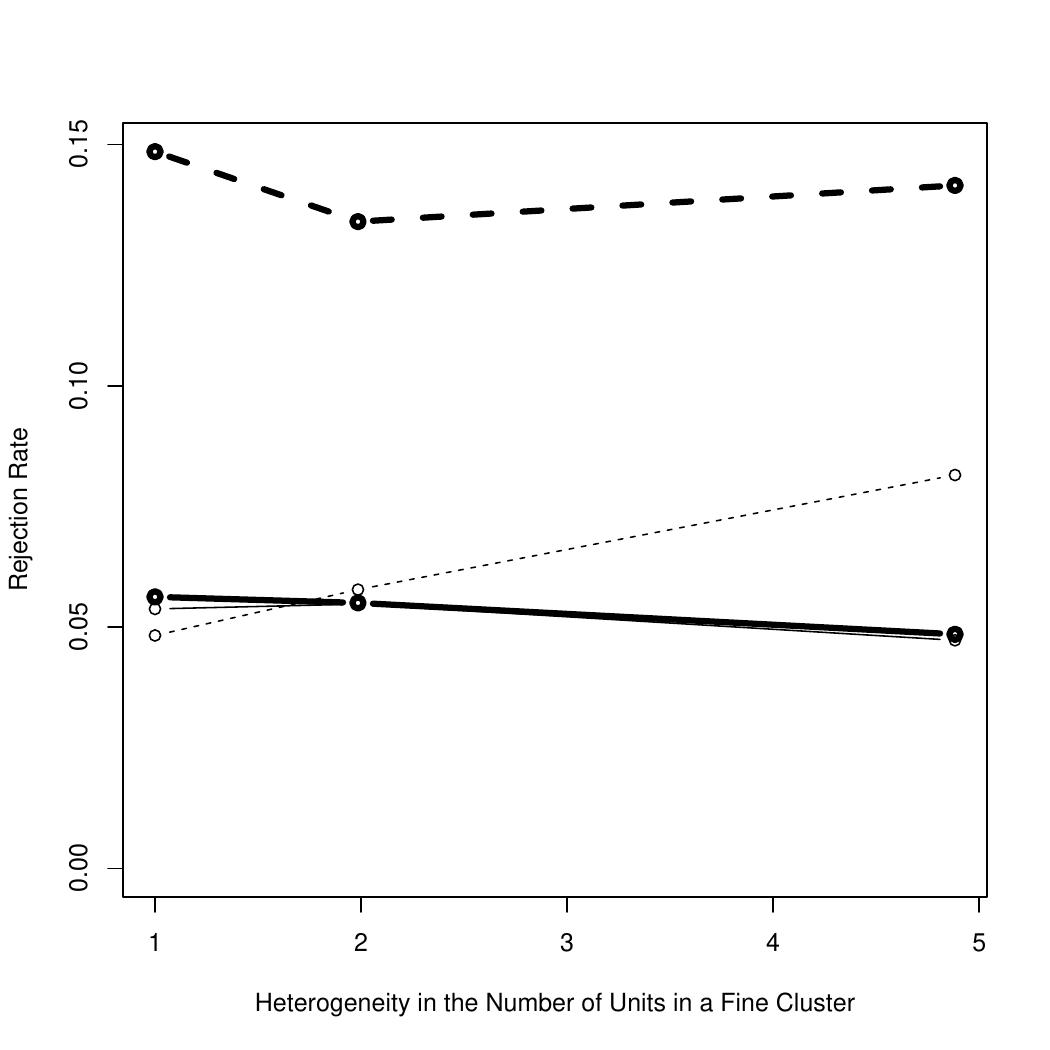}
    \subcaption{Size}
\end{minipage}
\begin{minipage}[b]{0.49\hsize}
    \centering
    \includegraphics[width=0.9\hsize]{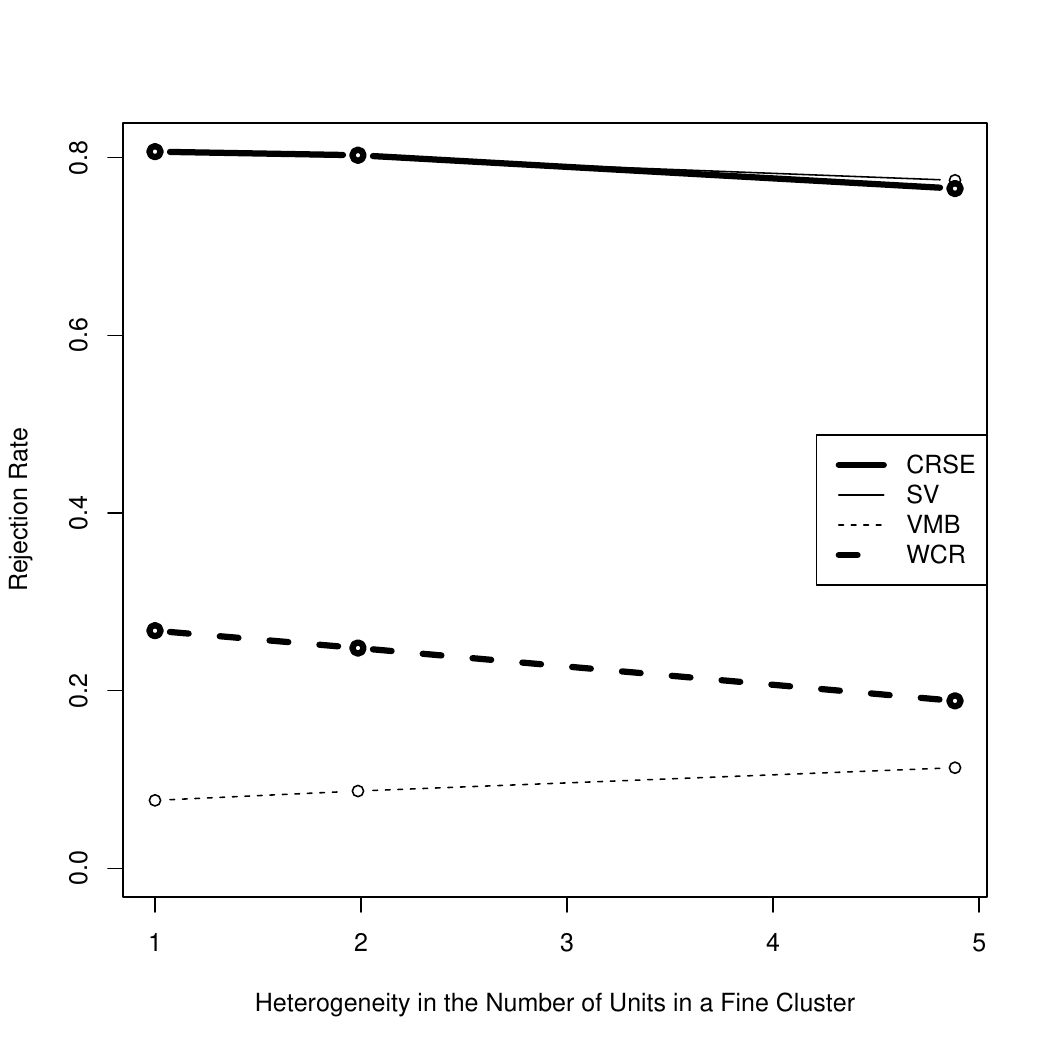}
    \subcaption{Power}
\end{minipage}
\caption{How the heterogeneity in the number of units in a fine cluster affects the performance of the four tests. I denote the rejection rate of the CRSE test by thick solid lines, that of the SV test by thin solid lines, that of the VMB test by thin dotted lines, and that of the WCR test by thick dotted lines. $\bar{g} = n_{g} = 12$ for all $g$. In each gross cluster, half of $n_{g}$ fine clusters have $n_{f} =100 + \Delta n$ units, and the other half have $n_{f} =100 - \Delta n$ units. $h_{f} \equiv \frac{100 + \Delta n}{100 - \Delta n} \approx 1, 2, 5$. $z = 4,000$. In the left panel, $\rho_{U} = 0$, while in the right panel, $\rho_{U} = 0.1$.}
\label{fig:hetero_n_f}
\end{figure}

I consider heterogeneity in cluster size. First, I look at heterogeneity in $n_{f}$. In each gross cluster, half of $n_{g}$ fine clusters have $n_{f} =100 + \Delta n$ units, and the other half have $n_{f} =100 - \Delta n$ units, where $\Delta n$ takes the values of $0, 33$, and $66$. Denote the heterogeneity parameter by $h_{f} \equiv \frac{100 + \Delta n}{100 - \Delta n}$, which takes the values of, roughly, $1, 2$, and $5$. $z = 4,000$. Other setups are the same as the baseline. Figure \ref{fig:hetero_n_f} shows the results. The horizontal axis indicates $h_{f}$. The performance of all but the VMB tests does not change so much as the heterogeneity increases; the rejection rate of the VMB test slightly increases as the heterogeneity increases.

\begin{figure}[!htb]
\centering
\begin{minipage}[b]{0.49\hsize}
    \centering
    \includegraphics[width=0.9\hsize]{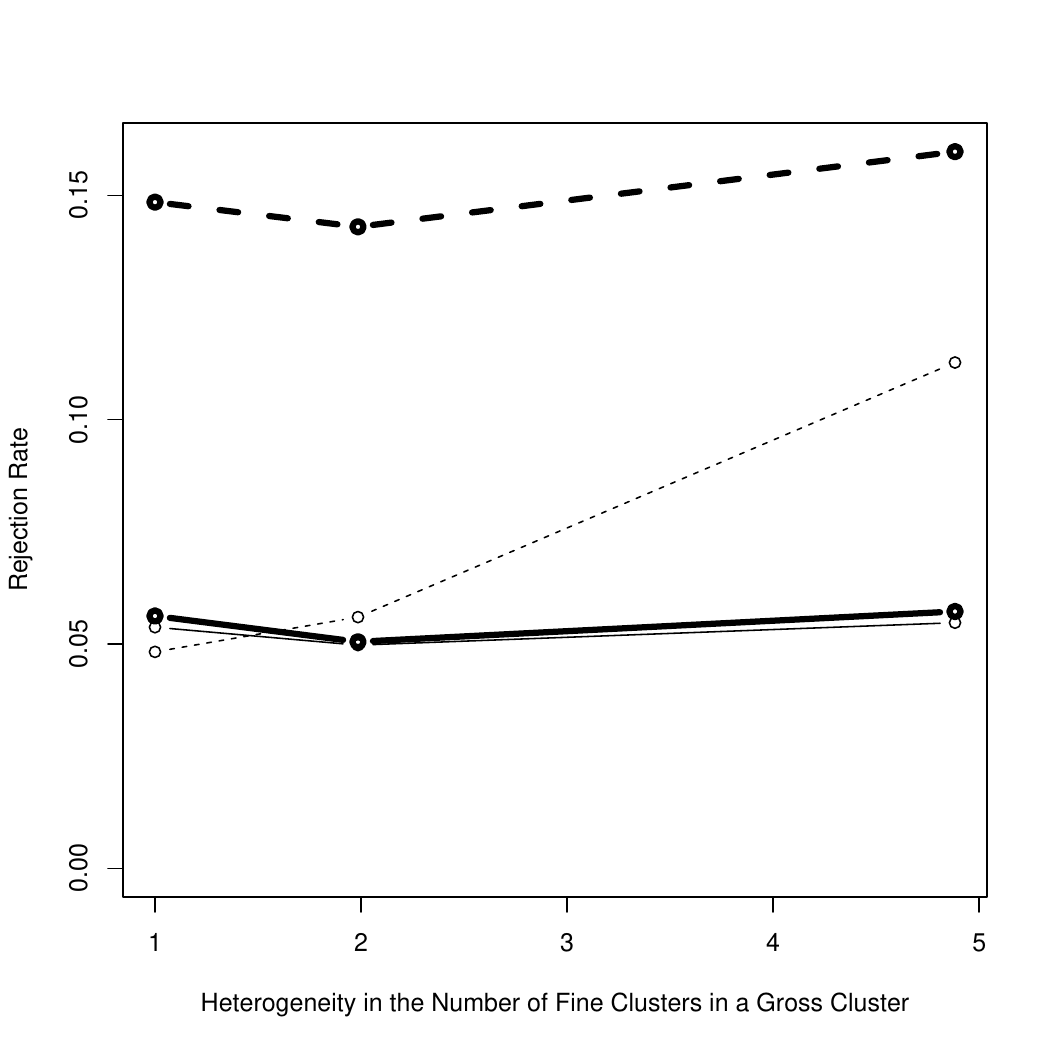}
    \subcaption{Size}
\end{minipage}
\begin{minipage}[b]{0.49\hsize}
    \centering
    \includegraphics[width=0.9\hsize]{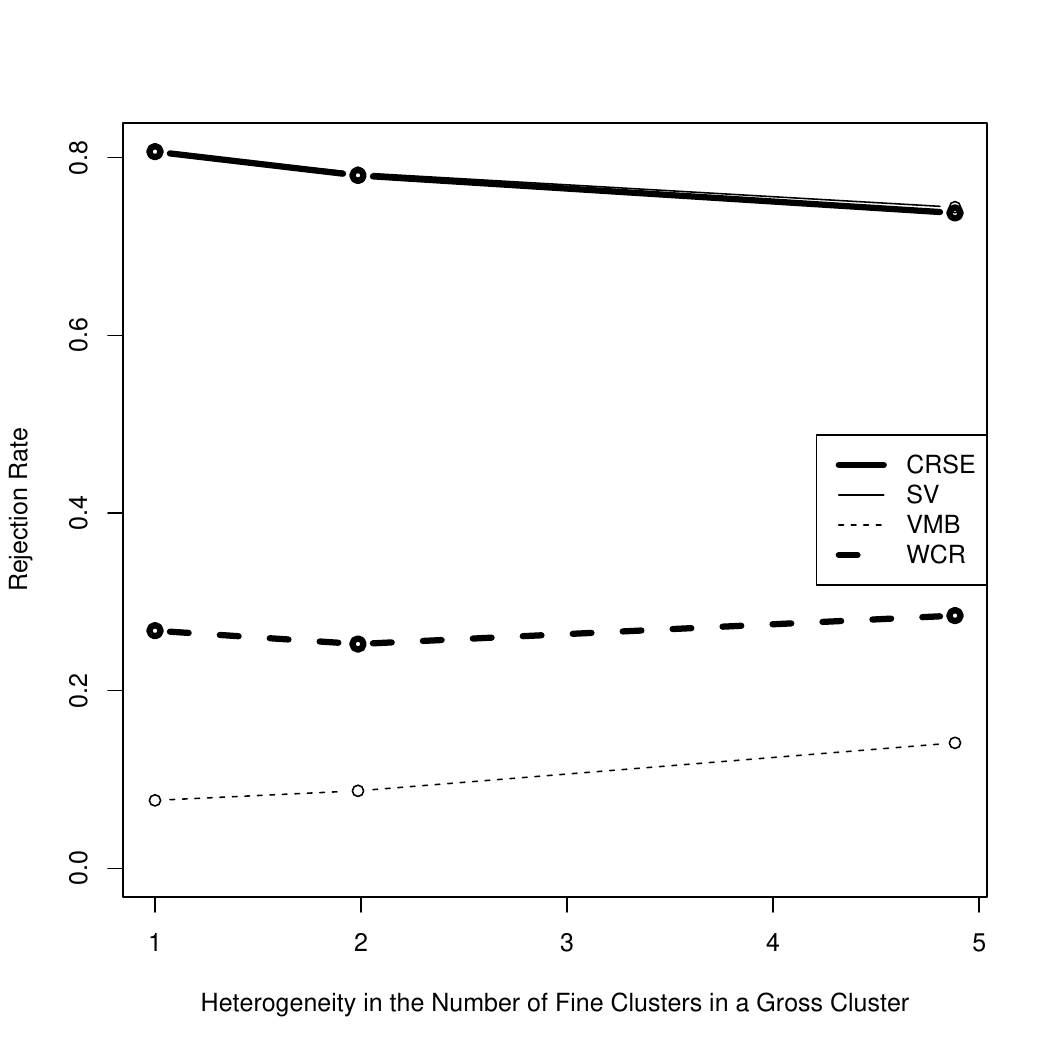}
    \subcaption{Power}
\end{minipage}
\caption{How the heterogeneity in the number of units in a fine cluster affects the performance of the four tests. I denote the rejection rate of the CRSE test by thick solid lines, that of the SV test by thin solid lines, that of the VMB test by thin dotted lines, and that of the WCR test by thick dotted lines. $\bar{g} = 12, n_{gf} = 100$ for all $g$ and $f$, respectively. Half of $\bar{g}$ gross clusters have $n_{g} = 12 + \Delta n$ fine clusters, and the other half have $n_{g} = 12 - \Delta n$ fine clusters. $h_{g} \equiv \frac{12 + \Delta n}{12 - \Delta n} \approx 1, 2, 5$. $z = 4,000$. In the left panel, $\rho_{U} = 0$, while in the right panel, $\rho_{U} = 0.1$.}
\label{fig:hetero_n_g}
\end{figure}

Next, I look at heterogeneity in $n_{g}$. Half of $\bar{g}$ gross clusters have $n_{g} = 12 + \Delta n$ fine cluster, and the other half have $n_{g} = 12 - \Delta n$ fine clusters, where $\Delta n$ takes the values of $0, 4$, and $8$. Denote the heterogeneity parameter by $h_{g} \equiv \frac{12 + \Delta n}{12 - \Delta n}$, which takes the values of, roughly, $1, 2$, and $5$. $z = 4,000$. Other setups are the same as the baseline. Figure \ref{fig:hetero_n_g} shows the results. The horizontal axis indicates $h_{g}$. The implications are the same as the previous figure's.

\subsubsection*{Very Small Sample}

One of the important differences between reclustering and previous methods is that previous methods rely on asymptotics and reclustering does not. 
In fact, \citet[2041--2042, Figures 1and 2]{MacKinnon_Nielsen_Webb_2023_SV} demonstrate in their simulation that the SV test does not work perfectly in small samples. Specifically, when $\bar{g}$ and/or $n_{g}$ are small ($\bar{g} < 24, n_{g} = 4, n_{gf} = 100$ and $\bar{g} = 8, n_{g} < 8, n_{gf} = 100$), the SV test slightly over-rejects. 
Accordingly, I examine the performance of the four tests for very small samples.
The ``minimum'' setup is $\bar{g} = n_{g} = 2, n_{gf} = 2$ for all $g$ and $f$, respectively, and $\rho_{U} = 0, z = 10,000$. 

\begin{figure}[!htb]
\centering
\begin{minipage}[b]{0.49\hsize}
    \centering
    \includegraphics[width=0.9\hsize]{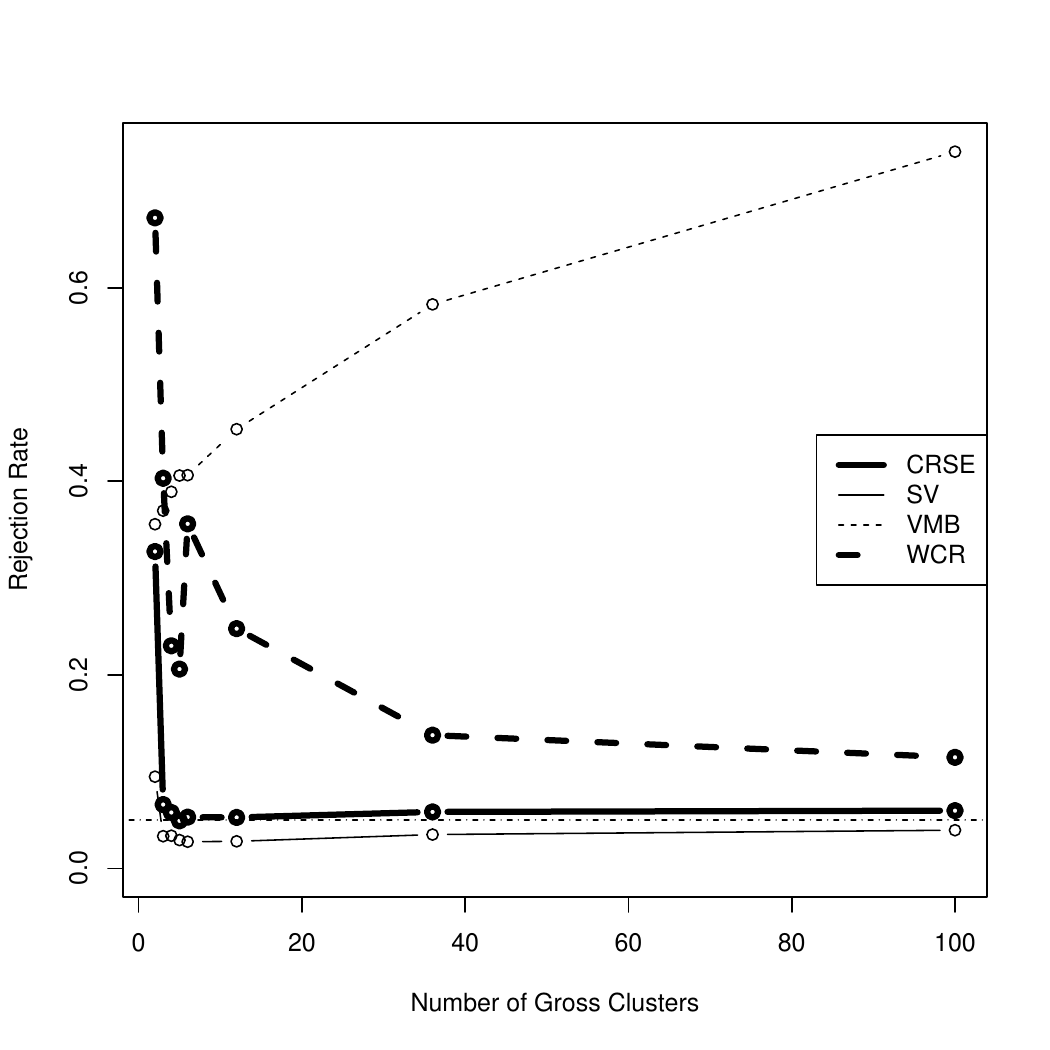}
    \subcaption{Number of Gross Clusters}
\end{minipage}
\begin{minipage}[b]{0.49\hsize}
    \centering
    \includegraphics[width=0.9\hsize]{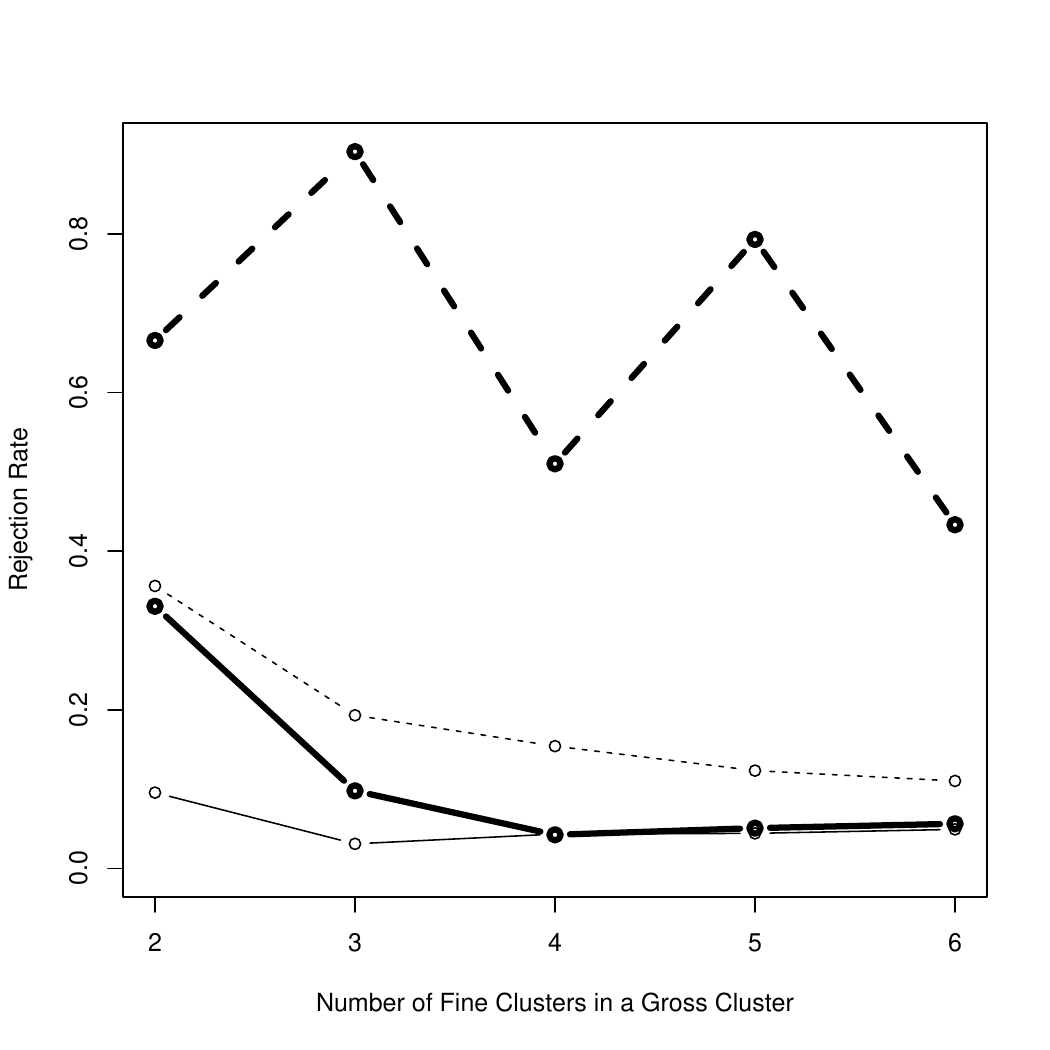}
    \subcaption{Number of Fine Clusters in a Gross Cluster}
\end{minipage}
\caption{The performance of the four tests in very small samples under Null Hypothesis 2. I denote the rejection rate of the CRSE test by thick solid lines, that of the SV test by thin solid lines, that of the VMB test by thin dotted lines, and that of the WCR test by thick dotted lines. $\rho_{U} = 0, z = 10,000$. In the left panel, $n_{g} = 2, n_{gf} = 2$ for all $g$ and $f$, respectively. In the right panel, $\bar{g} = 2, n_{gf} = 2$ for all $g$ and $f$, respectively.}
\label{fig:small}
\end{figure}

First, I change the number of gross clusters $\bar{g}$ from two to six by one, 12, 36, and 100. Other setups are the same as the minimum setup. 
The left panel of Figure \ref{fig:small} shows the results. The horizontal axis indicates $\bar{g}$. As I warned in the previous section, in the case of $\bar{g} \leq 3$, reclustering does not work. In fact, the rejection rate is close to $1/r^{*}$, that is, $1/3$ for $\bar{g} = 2$ and $1/15$ for $\bar{g} = 3$ (see also Table \ref{tab:bar_r_star}). 
Furthermore, for such a small sample, other tests do not work, either. Once $\bar{g}$ is not smaller than four, the CRSE test has a valid size. Nevertheless, the SV test slightly under-rejects, and the other two over-reject very much.

Second, I move the number of fine clusters per gross clusters $n_{g}$ from two to six by one. 
Other setups are the same as the minimum setup. 
The right panel of Figure \ref{fig:small} shows the results. The horizontal axis indicates $n_{g}$. 
In the case of $n_{g} \leq 3$, reclustering does not work. In fact, the rejection rate is close to $1/r^{*}$, that is, $1/3$ for $n_{g} = 2$ and $1/10$ for $n_{g} = 3$ (see also Table \ref{tab:bar_r_star}). 
Once $n_{g}$ is not smaller than four, the CRSE  and SV tests have a valid size, although the other two do not.\footnote{In the simulation by \citet[2042, Figure 2]{MacKinnon_Nielsen_Webb_2023_SV} ($\bar{g} = 8, n_{gf} = 100$), the SV test \textit{over}-rejects when $n_{g} = 3, 4$ but not when $n_{g} \geq 6$.} In the previous section, I argued that $n_{g} = 4$ is not large enough for reclustering to work, while there is no problem in this simulation.   

\begin{figure}[!htb]
\centering
\begin{minipage}[b]{0.49\hsize}
    \centering
    \includegraphics[width=0.9\hsize]{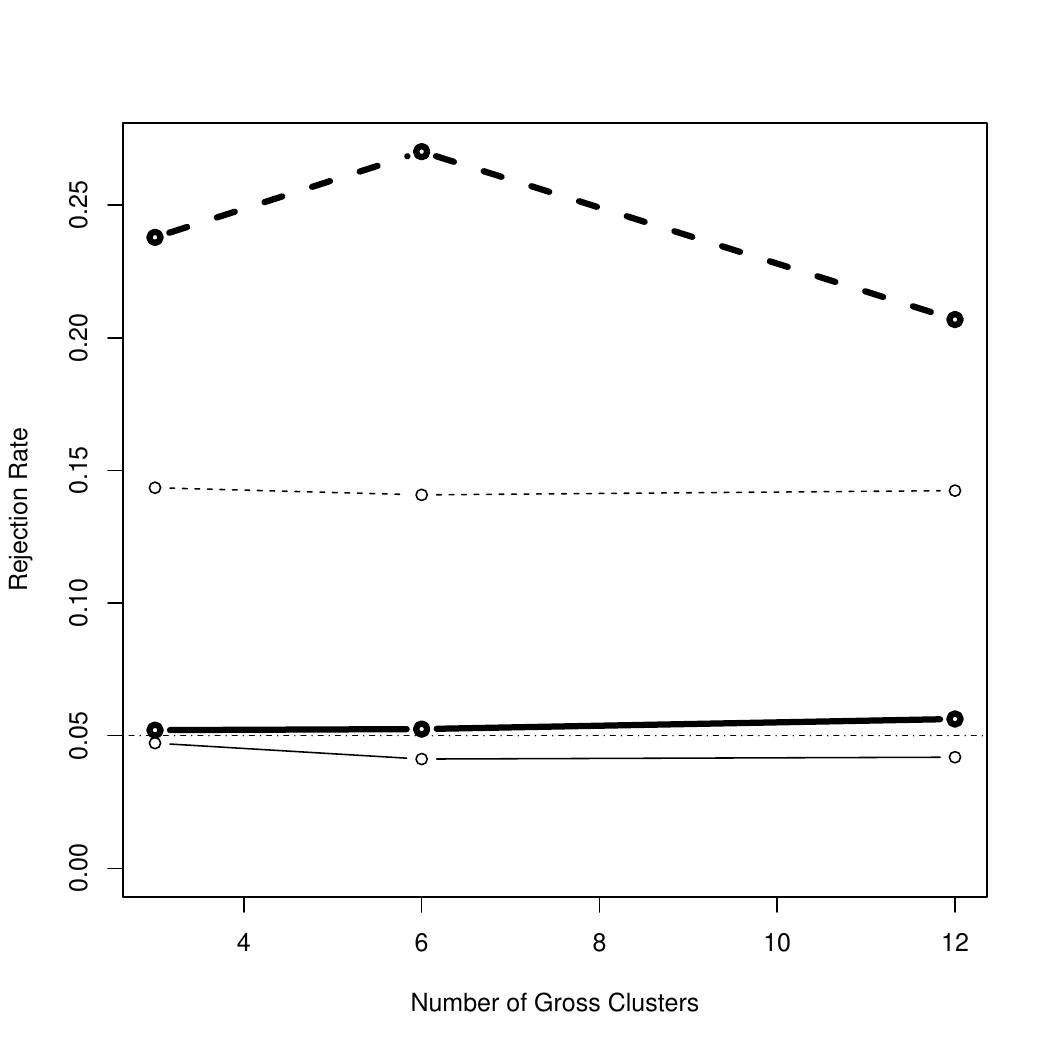}
    \subcaption{Number of Gross Clusters}
\end{minipage}
\begin{minipage}[b]{0.49\hsize}
    \centering
    \includegraphics[width=0.9\hsize]{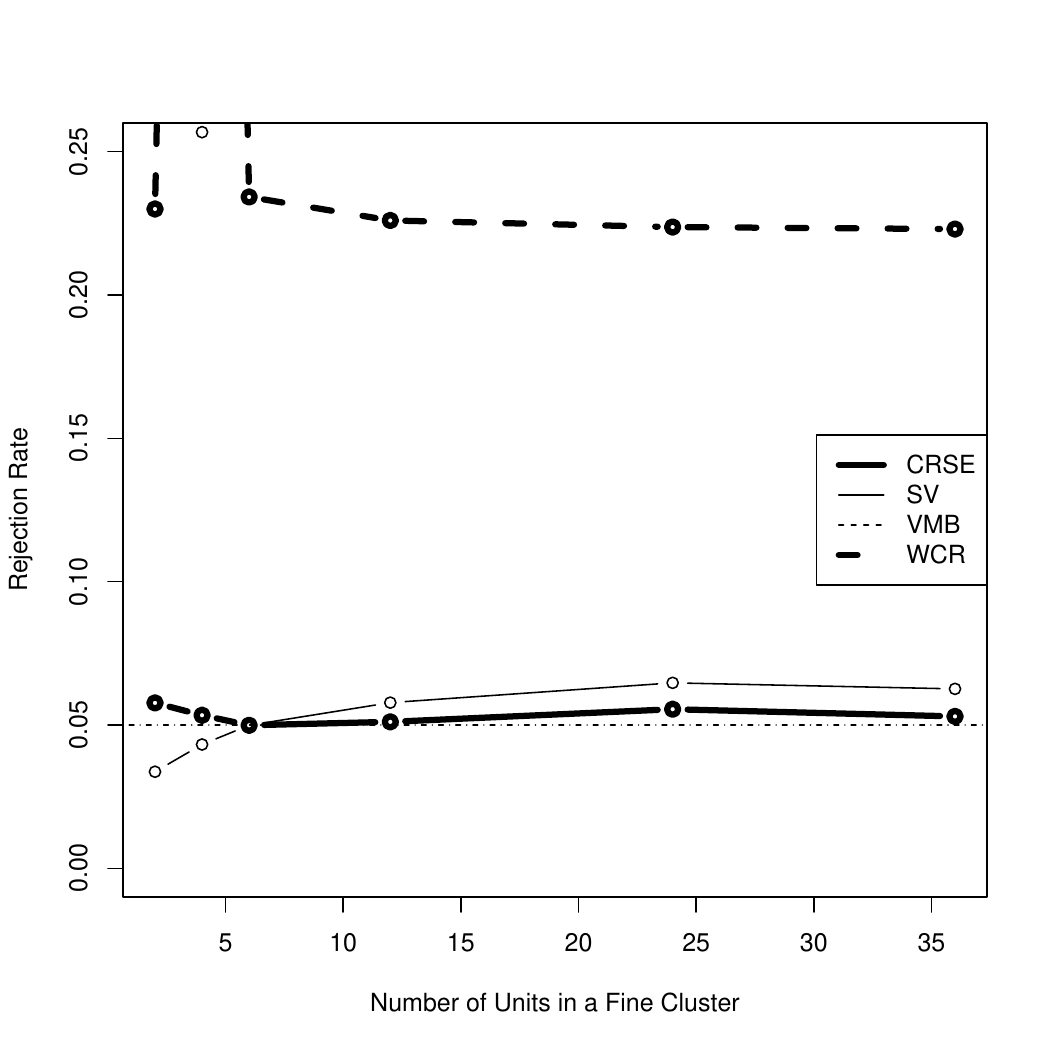}
    \subcaption{Number of Fine Clusters in a Gross Cluster}
\end{minipage}
\caption{The performance of the four tests in very small samples under Null Hypothesis 2. I denote the rejection rate of the CRSE test by thick solid lines, that of the SV test by thin solid lines, that of the VMB test by thin dotted lines, and that of the WCR test by thick dotted lines. $\rho_{U} = 0, z = 10,000$. In the left panel, $n_{g} = 4, n_{gf} = 2$ for all $g$ and $f$, respectively. In the right panel, $\bar{g} = 4, n_{g} = 2$ for all $g$ and $f$, respectively.}
\label{fig:small_2}
\end{figure}

Third, I increase $\bar{g}$ from 3 through 6 to 12, where $n_{g} = 4$. Other setups are the same as the minimum setup. The left panel of Figure \ref{fig:small_2} shows the results. Mostly, the CRSE test has a valid size, while the SV test under-rejects. The other two tests over-rejects extremely.

Finally, I increase $n_{gf}$ from 2, 4, 6, 12, 24, to 36, where $\bar{g} = 4$. Other setups are the same as the minimum setup. The right panel of Figure \ref{fig:small_2} shows the results. Mostly, the CRSE test has a valid size. In the cases of $n_{gf} < 6$, the SV test under-rejects, while in the cases of $n_{gf} > 6$, the SV test over-rejects. The other two tests over-rejects extremely.

To sum, when $\bar{g} + n_{g} \leq 5$, no tests work. When $n_{g} = 2$ and $\bar{g} \geq 4$, the CRSE test has a valid size, while the SV test does not. When $n_{g} = 4, n_{gf} = 2$, the CRSE test has a valid size, while the SV test uder-rejects. Otherwise, the CRSE and SV tests have nice performance, although the VMB and WCR tests do not.

\section*{APPLICATION\centering}

To show how to make much of reclustering, I reanalyze Study 2, foot ball games, of \citet{GHMM}.\footnote{The companion replication materials are \citet{GHMM_replication}, which exactly replicates not \citet{GHMM} but its earlier version \citet{GHMM_OSF}.} The study reanalyzes a seminal work, \citet{HMM}, which shows that when the college foot ball team in a county unexpectedly win within two weeks before the election, the incumbents in the presidential, gubernatorial, and senatorial races garner more vote share in the county. 

The unit of observation is county and race (original sample, 1985--2006). The dependent variable is the vote share of the incumbent party. The independent variable is a win of a foot ball team in the county within two weeks before the election minus the expected number of wins. I control lagged outcome, race dummies, year fixed effects, and county fixed effects. Following the ``respecified'' model of \citet{GHMM}, I do not include any other control variables.\footnote{In Figure 2 of \citet{GHMM}, my analysis corresponds to the top row (``Original sample''), the bottom half (``Respecified''), rightmost column (``Pooled''), and square (``Overall''). }

The quantity of interest is the coefficient of the independent variable. \citet{GHMM} employs the CRSE where the cluster is county. The left column of Table \ref{tab:application} reports the point estimate, the CRSE, and the $p$-value, which is barely above 0.05. This is consistent with \citet[Table L.3, Column 9]{GHMM_OSF}, while the corresponding value in \citet[Table L.3, Column 9]{GHMM} is less than 0.05. Anyway, the coefficient is nearly significantly positive.

\begin{table}[htb!]
\centering
\caption{The Coefficient Estimate of Unexpected Win.}
\label{tab:application}
\begin{tabular}{lrr}
  \hline
Cluster & County & State \\ 
  \hline
Estimate & 1.117 & 1.117 \\ 
  CRSE & 0.549 & 0.634 \\ 
  $p$ & 0.046 & 0.087 \\ 
   \hline
\end{tabular}

\end{table}

However, it is reasonable to cluster standard errors at state level instead of county level. The studied races are presidential, gubernatorial, and senatorial. The last two are explicitly contested statewide, while the the first race is also fought state by state due to electoral college system. In the right column of Table \ref{tab:application}, I present the CRSE where state is cluster. Now, the $p$-value surpass even 0.1. The numerical change is minor, although the tests come near leading to different conclusions.

\begin{figure}[!htb]
\centering
\begin{minipage}[b]{0.49\hsize}
    \centering
    \includegraphics[width=0.9\hsize]{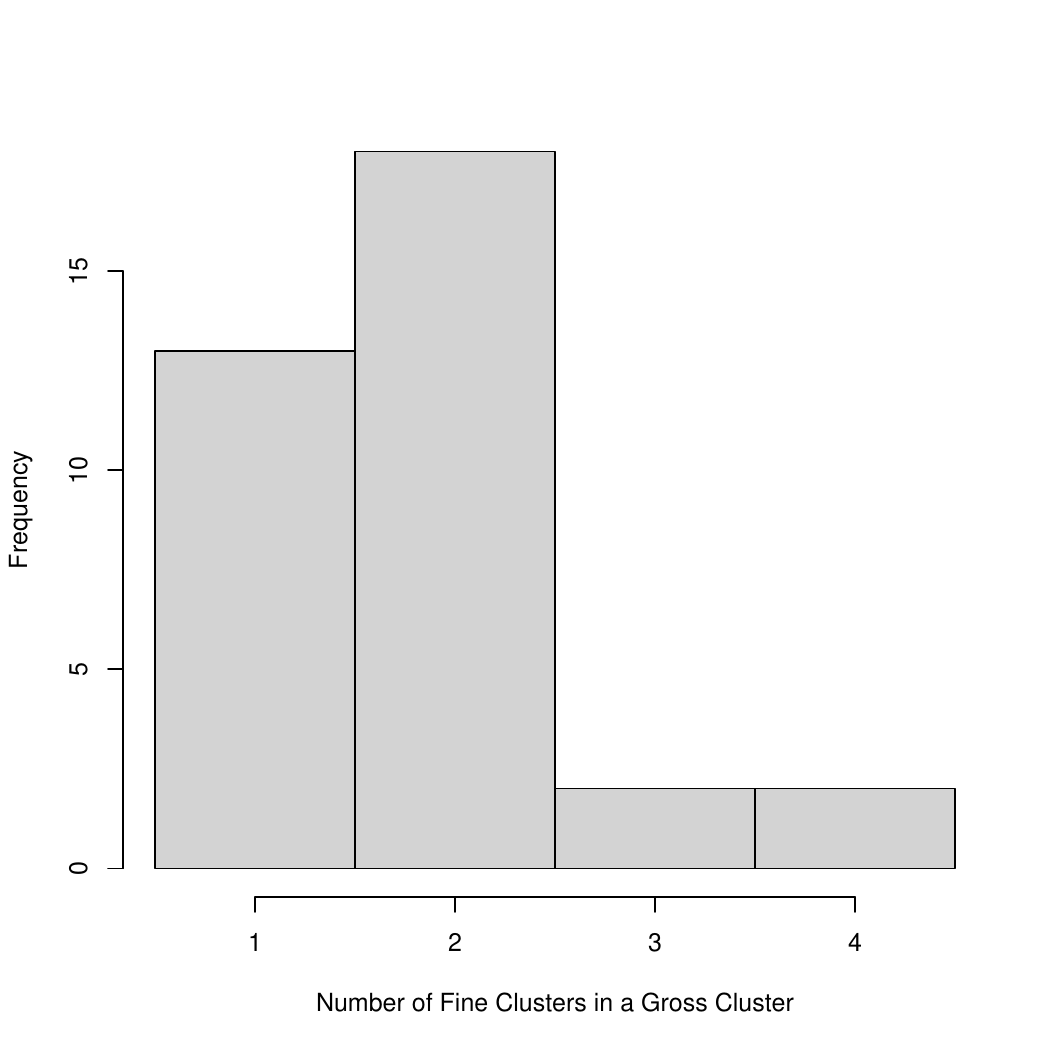}
    \subcaption{Number of Fine Clusters in a Gross Cluster}
\end{minipage}
\begin{minipage}[b]{0.49\hsize}
    \centering
    \includegraphics[width=0.9\hsize]{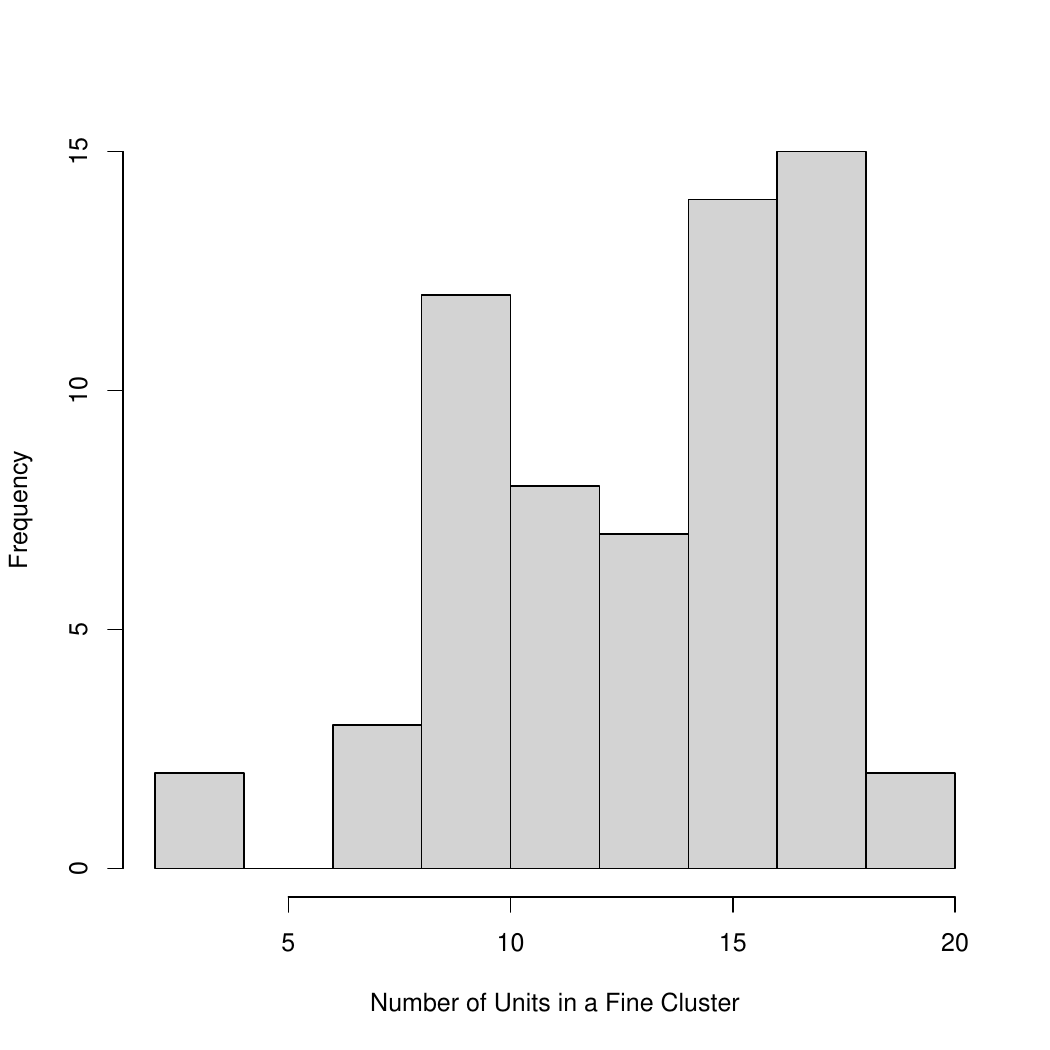}
    \subcaption{Number of Units in a Fine Cluster}
\end{minipage}
\caption{Cluster Size of \citet{GHMM}.}
\label{fig:application}
\end{figure}

Which level is more appropriate to cluster over? When analysts might be tempted to opt for the significant result, an empirical test is useful.
In this context, a fine cluster is a county, and a gross cluster is a state. There are $\bar{g} = 35$ states, $\bar{f} = 64$ counties, and $n = 852$ units. Figure \ref{fig:application} illustrates the distribution of $n_{g}$ (the number of counties in a state) and $n_{f}$ (the number of units in a county). They are heterogeneous in size.

\begin{figure}[!htb]
\centering
    \includegraphics[width=0.49\hsize]{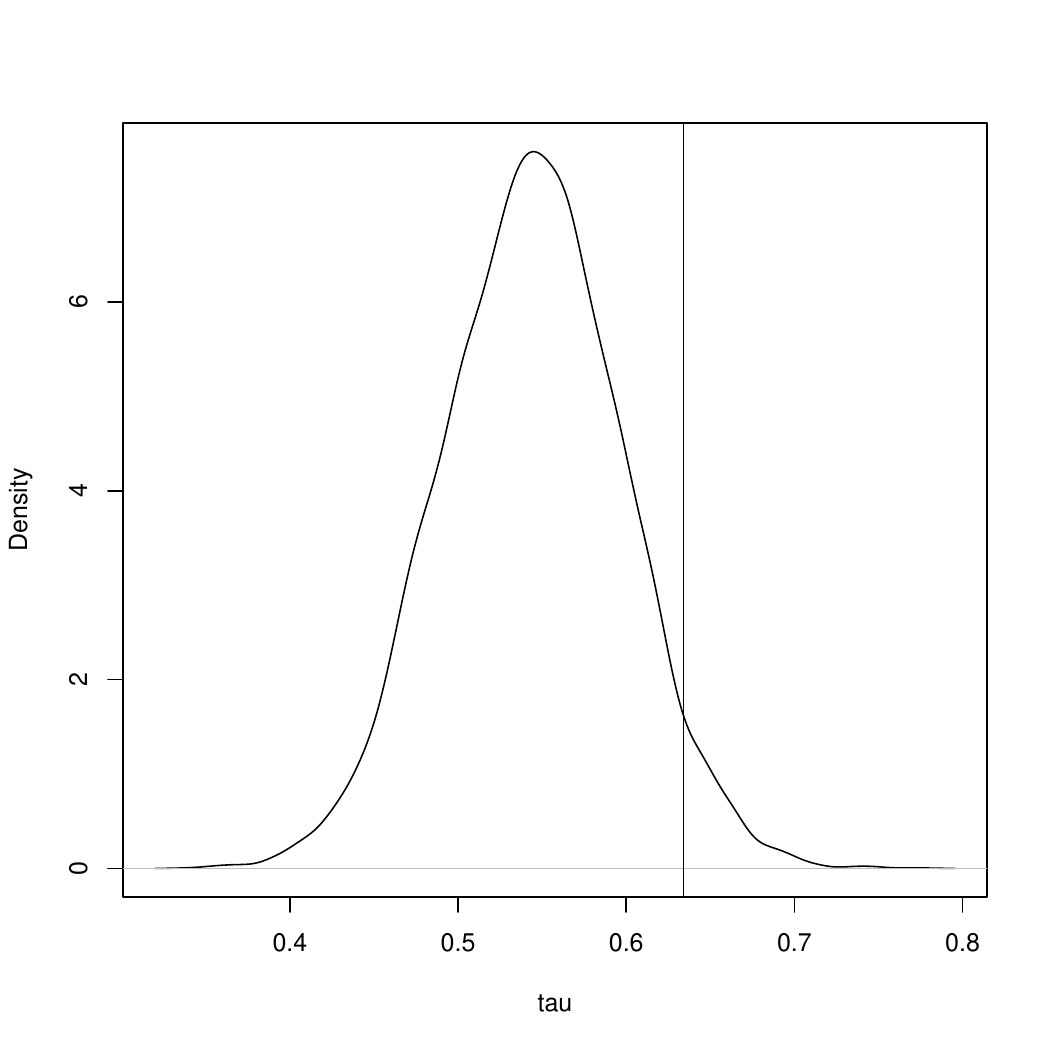}
\caption{The distribution of $\tau(\bm D, \bm g^{(r)})$ in the case of \citet[Table L.3, Column 9]{GHMM}. The vertical line indicates $\tau(\bm D, \bm g^{\mathrm{obs}})$.}
\label{fig:distribution_tau}
\end{figure}

I apply the four tests and report the $p$-values in Table \ref{tab:application_test}.\footnote{For reclustering, I permute counties $\bar{r} = 10,000$ times.} 
According to the CRSE test, the null hypothesis is rejected, although the three other tests fail to reject the null hypothesis. Taking into account the simulation results, I follow the CRSE test. 
Figure \ref{fig:distribution_tau} displays the distribution of $\tau(\bm D, \bm g^{(r)})$, where the vertical line indicates $\tau(\bm D, \bm g^{\mathrm{obs}}) = 0.660$. Obviously, the CRSE clustered at the observed state level is different from most of the CRSEs clustered at the reclustered state level. Therefore, the test rejects fine (county) cluster and supports the use of gross (state) cluster, where the original independent variable's coefficient is not significantly positive.

\begin{table}[htb!]
\centering
\caption{Tests for \citet[Table L.3, Column 9]{GHMM}.}
\label{tab:application_test}
\begin{tabular}{lc}
  \hline
Test & $p$ \\ 
  \hline
  CRSE &  0.038  \\ 
  SV & 0.044\\ 
  VMB & 0.695\\ 
  WCR & 0.575\\ 
   \hline
\end{tabular}

\end{table}

\section*{CONCLUSION\centering}

When there are fine and gross levels of clusters, at which level should we cluster the standard errors? This paper proposes a new method, reclustering, and evaluate its performance against a few previous tests by using Monte Carlo simulation.  
Overall, the CRSE and SV tests have almost the same size and power, while the VMB and WCR tests underperform. I also apply these tests to an application, \citet{GHMM}.

Reclustering has three merits.
First, it is intuitive. Under Null Hypothesis 2, gross clusters are meaningless. Thus, even if we reassign fine clusters to gross clusters, CRSE at gross cluster level should not change very much. If it does, we doubt and reject Null Hypothesis 2. 
Second, reclustering is easy to implement. Without any specific package, analysts should be able to permute fine clusters in any statistical software. 
Third, reclustering does not rely on asymptotics and does work in a very small sample (unless $\bar{g} + n_{g} \leq 5$).

The CRSE and SV tests exploit different information. 
The CRSE test considers score cross products ($\bm s_{gfi} \bm s'_{g'f'i'}$) that are fine cluster off-diagonal ($f \neq f'$). Among these cross products, the SV test limits to those that are gross cluster level diagonal ($g = g'$). Nevertheless, these two return similar results. The reason is a future research agenda.

This paper assumes that there are only two levels of clustering, while in practice, there may be more than two levels. In such situation, I suggest that researchers follow the sequential testing procedure by \citet[Section 3.3]{MacKinnon_Nielsen_Webb_2023_SV}. That is, they begin to test the finest level cluster against the second finest level cluster; if the null hypothesis is rejected, they proceed to test the second finest level cluster against the third finest level cluster, and so on. Some readers may be afraid of pretest bias. 
\citet[91]{Ibragimov_Muller_2016} recommend that they ``report the significance of results based on various clustering assumptions and interpret the resulting inference conditional on the validity of these assumptions.''

This paper studies the common, but restricted, setup of clustering. In future research, I plan to consider clustering in a more general context.

\clearpage

\bibliographystyle{apsr_fs}
\bibliography{ref}

\end{document}